\pdfoutput=1
\documentclass[pdflatex,sn-nature,iicol,super]{sn-jnl}
\usepackage{graphicx}
\usepackage{multirow}
\usepackage{amsmath,amssymb,amsfonts}
\usepackage{amsthm}
\usepackage{mathrsfs}
\usepackage[title]{appendix}
\usepackage{xcolor}
\usepackage{textcomp}
\usepackage{manyfoot}
\usepackage{booktabs}
\usepackage{algorithm}
\usepackage{algorithmicx}
\usepackage{algpseudocode}
\usepackage{listings}
\usepackage{comment}
\usepackage{geometry}
\usepackage{float}
\usepackage{subcaption}
\usepackage[export]{adjustbox}
\theoremstyle{thmstyleone}

\usepackage{float}
\usepackage[export]{adjustbox}

\definecolor{pdcolor}{rgb}{1,0.5,0}

\definecolor{pdblue}{rgb}{0,0,1}

\definecolor{rkgreen}{rgb}{0,1,0}

\begin{document}
\title{Active movement of foraging sea turtles generates anomalous
  looping}

\author[1]{\fnm{Vijay} \sur{Kumar}}\email{vksms149@gmail.com}
\author[1]{\fnm{Vladimir V.} \sur{Palyulin}}\email{v.palyulin@gmail.com}
\author[2]{\fnm{Perla} \sur{Roman-Torres}}\email{perlaromantorres@gmail.com}
\author[2]{\fnm{Christophe} \sur{Eizaguirre}}\email{c.eizaguirre@qmul.ac.uk}
\author[3,4]{\fnm{Rainer} \sur{Klages}}\email{r.klages@qmul.ac.uk}
\affil[1]{\orgname{Skolkovo Institute of Science and Technology}, \state{Moscow}, \country{Russia}}
\affil[2]{\orgdiv{School of Biological and Behavioural Sciences}, \orgname{Queen Mary University of London}, \country{UK}}
\affil[3]{\orgdiv{Centre for Complex Systems, School of Mathematical Sciences}, \orgname{Queen Mary University of London},\country{UK}}
\affil[4]{\orgname{London Mathematical Laboratory}, \country{UK}}

\abstract{Animals inhabiting diverse environments by moving across different spatial scales, from insects \cite{KaSh83,LCK13}
  to birds
  \cite{Vis96,VCTBN25},
  marine predators
  \cite{Sims08,BPA23},
  mammals \cite{RMM04,MCS23}
  and
  even humans \cite{BHG06,GHB08},
  often display apparently random
  movement paths. Over the past decade, novel biologging technologies have recorded these patterns in increasing detail, generating a wealth of
  experimental data \cite{Kays15,Nath22}.
  A central challenge is to understand such complex patterns by
  constructing data-driven mathematical models. Many animal movements
  depart from Brownian motion, as described by correlated random walks
  \cite{CPB08},
  L\'evy walks,
  \cite{VLRS11}
  or active particle dynamics
  \cite{RBELS12}.
  Yet, these movement models do not incorporate long-term
  non-Markovian memory extracted from experimental trajectories
  \cite{LICCK12,Diet22}.
  Here, we construct a stochastic generalised Langevin equation from
  satellite tracking data for loggerhead sea turtles {\em (Caretta
    caretta)} foraging off the coast of West Africa. We find that
  these turtles exhibit active movement characterised by large-scale
  loops that are not explained by ocean currents \cite{LLL06} or
  chirality \cite{BeDiL16}.
   These loops maintain movement within a specific foraging region
and, over intermediate timescales, generate superdiffusion similar to
L\'evy walks. We thus identify a loop-based form of active anomalous
search related to foraging patterns observed across a wide range of
animal species \cite{CZR03,LCK13,MCS23,VCTBN25}, which
may inspire robotic search strategies \cite{GKV25}
  and AI-based metaheuristic optimisation algorithms \cite{ChaDu18}.}

\maketitle

For almost a century, the apparently random movement patterns of
organisms, including cells, have been modelled using simple stochastic
processes such as random walks \cite{Ross05,Pea06}
or Langevin
equations \cite{PPL17,Kla24}.
Roughly three decades ago,
it became clear that such movements cannot always be explained
by Brownian motion, which predicts a mean-squared
displacement (MSD) that grows linearly in time, $\langle
x^2\rangle\sim t^{\alpha}$ with $\alpha=1$. Many experiments instead revealed a nonlinear temporal scaling of the MSD
\cite{VLRS11,ZDK15} with
$\alpha\neq1$, which is the hallmark of {\em Anomalous Diffusion} (AD)
\cite{MeKl00,MJCB14},
a phenomenon that occurs across various fields of science
\cite{KRS08,WaKo23}.
Anomalous biological motion has since then been modelled using
advanced stochastic processes like L\'evy walks \cite{VLRS11,ZDK15},
continuous time random walks \cite{MeKl00,VOCGTNA21}, and generalised
Langevin equations with memory kernels \cite{LICCK12,MSDRN20}.

The concept of Brownian motion was originally developed to describe
the apparently random dynamics of tracer particles in a fluid
passively driven by molecular collisions. Organisms, however, are
self-propelled, due to energy uptake from the environment by
converting it into persistent motion. The resulting self-generated
movement is described by generalisations of ordinary Langevin dynamics
known as {\em Active Particle} (AP) models \cite{BeDiL16,RBELS12}.
This theory is widely applied to study the collective dynamics of
self-driven agents with the emergence of active
matter \cite{Rama10,MJR13}.
At the level of individual organisms, movement also depends on
biomechanics, internal states, perception and interaction with the
environment. Revealing the interplay between these processes in a
given ecosystem is central to {\em Movement Ecology} (ME)
\cite{Nath08,GiMa26}, which
has developed a mathematical framework for explaining movement
tracking data using state-space models \cite{PPL17,ANC21}, hidden
Markov processes \cite{PPL17,GAL23} and (correlated) random walks
\cite{CPB08,MCB14}.

Here we show that understanding the foraging movements of loggerhead
sea turtles naturally connects the fields of ME, AP and AD
\cite{KLK26}. By extracting velocity probability distributions and
velocity autocorrelation functions from satellite tracking data, we
find that crucial information is contained in the turtles'
autocorrelations, yielding transient anomalous superdiffusion with an
exponent $\alpha>1$ up to intermediate time scales. These non-trivial
autocorrelations represent a form of biological activity that
generalises a widely used active particle model with exponentially
correlated velocities \cite{KMDL14,Sza14}.  The resulting active
anomalous movement remains confined within a foraging region shaped by
environmental boundaries through the formation of large-scale looping.
We show that these features can be consistently reproduced by a
conceptually simple, data-driven generalised Langevin equation
\cite{Diet22}, providing a coarse-grained stochastic model of turtle
movement. All main results have been confirmed by analysing the
tracking data of ten different turtles. Our modelling framework
complements the biologically motivated analysis in Ref.~\cite{RSS25},
which assesses how habitat-specific environmental characteristics are
associated with turtle movement.

\section*{Results} \label{sec:res}
\subsection*{Experimental data analysis}
In total, we analysed movement data from ten loggerhead sea turtles
nesting at the Cabo Verde archipelago and foraging off the West
African coast \cite{RSS25}. Here we show representative results for
one female loggerhead sea turtle, Mokamba, which was tracked from the
island of Fogo for 399 days in 2012/13. Details for all ten individual
turtles, tracking data and analytical procedures are provided in the
Methods section and the Supplementary Information (SI). Although
individual trajectories differed, the main statistical features we
report below were observed across all ten turtles (SI Sec.1).

\begin{figure}[htb]

  \vspace*{-0.5cm}
    \centering
      \includegraphics[width=1.0\linewidth]{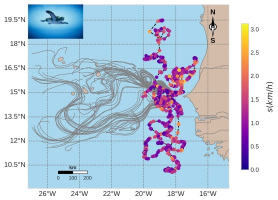} 
    \caption{{\bf Sea turtle foraging vs.\ tracer motion.} Path of a
      loggerhead sea turtle during its foraging phase off West Africa,
      after nesting at the Cabo Verde archipelago. The 1078 data
      points are coloured by speed, highlighting intermittent
      fluctuations between slower and faster movement. Grey
      trajectories show 50 simulated Lagrangian tracer particles
      released near the centre of the turtle's foraging region. The
      passive tracers disperse differently from the turtle path, which
      remains bounded within a specific foraging region by displaying
      distinct large-scale loops..}
    \label{fig:Fig1}

   \vspace*{-0.5cm}
\end{figure}

\begin{figure*}[htb]
    \centering
     \includegraphics[width=1.0\linewidth]{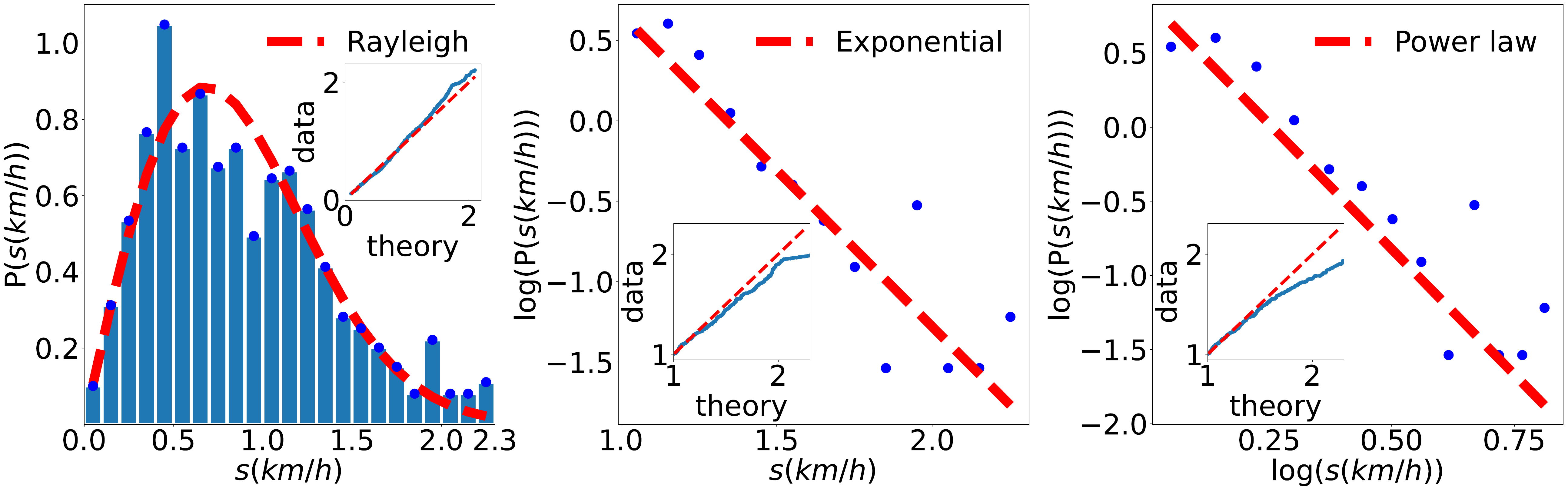} 
    \caption{{\bf Speed distribution function.}  Left: The speed
      distribution along the turtle path in Fig.~\ref{fig:Fig1} is
      well described by a Rayleigh functional form near the
      centre. Middle, Right: Semi-logarithmic and double-logarithmic
      plots indicate that the tail is better represented by an
      exponential or power law than by the Rayleigh
      distribution. Insets: Quantile-quantile plots show that the
      Rayleigh form reproduces the centre, whereas the tail decays
      approximately exponentially.}
    \label{fig:Fig2}
\end{figure*}
Figure~\ref{fig:Fig1} shows the foraging path of Mokamba, which
consists of longitude and latitude coordinates recorded at irregular
time intervals regularised by linear interpolation. Simulated
Lagrangian tracer particles released in the same region diverge from
the turtle trajectory (see SI Sec.2.5 for details), demonstrating that
the observed path was not simply generated by ocean currents
\cite{LLL06}. Instead, Mokamba displays bounded movement structured by
distinct large-scale loops interspersed with intermittent changes in
speed by alternating between slower and faster movement.  This
movement is spatially asymmetric, as it is bounded longitudinally by
the coastline to the east and deeper ocean waters to the west, while
extending over a broader latitudinal range. Similar results for the
other nine turtles can be found in SI Sec.1.1.

After data cleaning (SI Sec.1.3) we extracted the speed distribution
for Mokamba displayed in Fig.~\ref{fig:Fig2}. At lower and intermediate
speeds, the distribution is well fitted by a Rayleigh distribution,
whereas the tail decays exponentially. This pattern is consistent with
the velocity distributions along both longitudes and latitudes, which
display Gaussian centres and exponential tails (SI Sec 1.5). These
different functional forms for lower and higher speeds match the
intermittent character of the dynamics observed in
Fig.~\ref{fig:Fig1}. Biologically, such intermittency may reflect
alternation between slower area-restricted movements and faster
relocation phases, searching for food. Analogous results for the other
nine turtles are available in SI Sec.1.4.

\begin{figure}[b]
    \centering
     \includegraphics[width=1.0\linewidth]{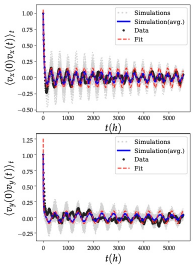} 
    \caption {{\bf Velocity autocorrelation functions.} Top: Temporal
      decay of the longitudinal velocity autocorrelation function
      $\langle v_{x}(0)v_{x}(t)\rangle_{t}$, computed as a time
      average along the turtle path in Fig.~\ref{fig:Fig1}. Diffferent
      lines correspond to the estimate and modelling approaches
      indicated in the legend and described in the text. The
      autocorrelation decays rapidly at short times and is followed by
      persistent regular oscillations at longer times, with
      approximately constant frequency and amplitude. Bottom:
      Corresponding latitudinal velocity correlation function,
      $\langle v_{y}(0)v_{y}(t)\rangle_{t}$. While oscillations are
      present, there are less pronounced and regular than in the
      longitudinal direction.}
    \label{fig:Fig3}
\end{figure}
Figure~\ref{fig:Fig3} displays the velocity autocorrelation functions
(VACFs) for the velocity components $v_x$ and $v_y$ along longitudes
($x$-axis) and latitudes ($y$-axis), respectively. They were both best
fitted with $f(t) = A \exp(-Bt) + C \cos(Dt)$, where $A,B,C,D$ are
real constants. That both VACFs decay exponentially on short time
scales is supported by additional analyses (SI Sec 1.8). Our most
striking result is the existence of sustained oscillations in the
longitudinal VACF, with approximately constant frequency and amplitude
over the analysed time window. Persistent oscillations in the
latitudinal VACF are also present but less regular (SI Sec.1.9). This
directional difference is consistent with the spatial asymmetry of the
turtle motion observed in Fig.~\ref{fig:Fig1}: Movement is more
strongly bounded along longitudes, whereas the foraging region is more
open along latitudes. The persistence of the oscillations,
particularly in the longitudinal direction, suggests that they reflect
structured movement actively generated by the turtle itself, on top of
exponentially decaying short-term noise; for results of the other nine
turtles see SI Sec.1.9.

Turning-angle distributions are symmetric around zero for all turtles
(SI Sec.1.8), ruling out chirality in the dynamics potentially generated internally by an animal
\cite{BeDiL16}. Similarly, we found no significant cross-correlations between
$v_x$ and $v_y$ (SI Sec.1.10), excluding effects such as odd
diffusivity \cite{HEM21}. As shown below, these oscillatory VACFs
provide a statistical signature of the large-scale looping visible in
Fig.~\ref{fig:Fig1}.
\begin{figure}[t]
    \centering
     \includegraphics[width=1.0\linewidth]{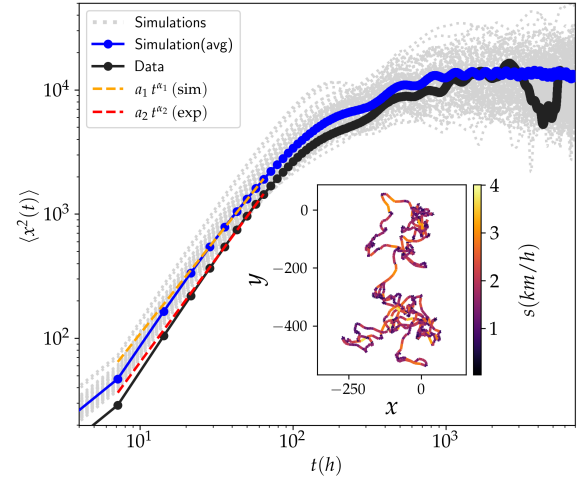} 
    \caption{{\bf Mean-squared displacement}. Mean-squared
      displacement computed as a time average along the trajectory in
      Fig.~\ref{fig:Fig1}, shown together with fits and simulation
      results. The turtle displays superdiffusive movement over
      intermediate time scales, with an exponent of $\alpha_2\simeq
      1.68$, before crossing over towards localisation at longer
      times. The inset yields a representative trajectory generated by
      the data-driven stochastic model, displaying large-scale loops
      and intermittent speed fluctuations qualitatively similar to
      those observed in Fig.~\ref{fig:Fig1}}
    \label{fig:Fig4}
\end{figure}
Mokamba's MSD along $x$ is shown in Fig.~\ref{fig:Fig4}. A fit to
the data reveals superdiffusive movement over intermediate timescales,
up to six days, followed by a transiently subdiffusive regime that
precedes localisation at longer times. Similar results for the MSD
along $y$, and analogous results for all the other turtles, are
presented in the SI Sec.1.7.

We next examined how the observed movement structure couples with
environmental conditions. The foraging regions used by Mokamba
(Fig.~\ref{fig:Fig1}) and the other nine turtles (SI Sec.1.1) are
bounded longitudinally by the coastline on one side and by deep
offshore waters on the other. This spatial confinement is consistent
with habitat-specific environmental gradients in the region, including
bathymetry, sea surface temperature and the chlorophyll-a
concentration as a proxy for food availability, where the latter two
quantities are negatively correlated \cite{RSS25}. The Canary current
may limit further excursions along latitudes. From
Fig.~\ref{fig:Fig3}, the average half period of the oscillations
along $x$ is $T_{x,1/2}\simeq165\,h$ to $204\,h$. According to the MSD
in Fig.~\ref{fig:Fig4}, this corresponds to a mean displacement of
approximately $\sqrt{\langle x^2(T_{x,1/2})\rangle_T}\simeq 84\,km$ to
$100\,km$ (see SI Sec.3 for details). An average longitudinal
extension of the foraging region in Fig.~\ref{fig:Fig1} is difficult
to determine, as it strongly fluctuates along latitudes. However, a
rough measure of $\Delta x\simeq 100\,km$ is of the same order of
magnitude as the extracted mean displacement, suggesting the loops of
Mokamba are well adjusted to cover the used foraging region on a
sufficiently large scale. A similar scale matching has been confirmed
for four other turtles (SI Sec.3).

At finer spatial scales, turtle movement may be associated with
chlorophyll-a concentration gradients, as chlorophyll is a known proxy
for ocean productivity and potential prey fields, rather than a direct
measure of prey availability. Generally turtles spend proportionally
less time in regions of high chlorophyll-a concentration, which
correspond to areas of high species richness, including turtle
predators. Instead, in this region of the world, turtles tend to swim
perpendicular to the chlorophyll gradient in the direction of
intermediate concentrations and associated intermediate sea surface
temperatures \cite{RSS25}. We found no evidence that day and night
generates different velocity distributions (SI Sec.1.11). As indicated
above already, this complete set of analyses was rolled out for all
ten turtles. Although individual differences were pronounced, the main
results reported above, i.e., the existence of large-scale loops,
speed distributions with Gaussian centres and exponential tails,
oscillatory VACFs and a superdiffusive MSD, were confirmed for all
turtles (SI Sec.1).

\section*{Data-driven stochastic modelling}
Using these empirical results, we construct a data-driven stochastic
model that reproduces the coarse-grained dynamics of the turtle
movement. Our model is inspired by the generalised Langevin dynamics
used to describe bumblebee flights \cite{LICCK12,LCK13} and cell
migration \cite{Diet22,Kla24}. The simplest version directly driven by
data consists of overdamped generalised Langevin dynamics in the form
of $d{\bf x}/dt={\bf F}({\bf x})+\boldsymbol{\zeta}(t)$, where ${\bf
  x}$ denotes the position vector in the plane, ${\bf F}({\bf x})$ an
external force and $\boldsymbol{\zeta}(t)$ a random variable sampled
from a velocity probability distribution that can be
correlated. Interestingly, this equation also defines an active
particle model \cite{KLK26} in which activity is encoded by
non-trivial correlation decay \cite{KMDL14,Sza14}.  If
$\boldsymbol{\zeta}(t)$ is chosen as exponentially correlated Gaussian
noise, the model reduces to the widely studied active
Ornstein-Uhlenbeck particle \cite{MBC21}. Our data analysis suggests
using, as model input, the velocity distributions extracted for
Mokamba (SI Sec.2), corresponding to the speed distribution in
Fig.~\ref{fig:Fig2}, together with the VACFs shown in
Fig.~\ref{fig:Fig3}. To simplify our model for simulations, we
neglected the exponential tails and used only fitted pure Gaussians
for both velocity components. These Gaussian variables were correlated
according to the fitted VACF functions displayed in
Fig.~\ref{fig:Fig3}. To account for the large-scale longitudinal
confinement of the foraging region, the model was amended with an
external force applied in the $x$ direction only, representing the
effective boundary conditions (see Methods and SI Sec.2.1-2.3 for
details). The numerical results obtained from this stochastic model
are presented in Figs.~\ref{fig:Fig3},\ref{fig:Fig4}.
The bold blue line in Fig.~\ref{fig:Fig3} shows the ensemble-averaged
VACF from simulations, while the thin grey lines yield time-averaged
VACFs from individual simulated trajectories. This representation
illustrates the variability expected among finite stochastic
realisations of the model. Similarly, Fig.~\ref{fig:Fig4} compares
the empirical MSD with simulated MSDs and their ensemble average. A
fit to the simulated ensemble average gives an exponent of
$\alpha_1\simeq 1.55 $, in agreement with the experimental scaling.
The inset of Fig.~\ref{fig:Fig4} depicts a representative
trajectory generated by the stochastic model. Despite the
simplification of neglecting exponential velocity tails, the simulated
path well reproduces the main qualitative features of the empirical
trajectory in Fig.~\ref{fig:Fig1}, including intermittent speed
fluctuations, large-scale looping and eventual localisation. At finer
scales, the simulated path appears more 'rugged' than the empirical
trajectory, which may reflect the overdamped approximation with
omission of acceleration. Overall, the excellent agreement between
simulated data and empirical results demonstrates that oscillatory
VACFs, combined with effective spatial confinement, are sufficient to
generate the large-scale looping visible in Fig.~\ref{fig:Fig3} and
the observed transition towards localisation. Respective results for
four other turtles are shown in SI Sec.2.4.

\section*{Discussion}
In conclusion, by statistically analysing satellite-tracking movement
data of ten loggerhead sea turtles, with Mokamba used as a
representative example (see the SI for analogous results of the other
nine turtles), we identified a recurrent movement mode that we call
{\em anomalous looping}. Here, the term anomalous refers to the
transient superdiffusive dynamics generated by the turtles over
timescales of several days. In distinct contrast to L\'evy walks,
whose hallmark is a power-law distribution of jump lengths
\cite{VLRS11,ZDK15}, the superdiffusion observed here emerges from the
specific long-term temporal memory in the dynamics, as revealed by
oscillatory VACFs. The corresponding velocity distributions are not
L\'evy but instead exhibit Gaussian centres with exponential
tails. These findings highlight the importance of extracting
correlation functions from experimental data, rather than inferring
movement mechanisms from probability distributions alone
\cite{LICCK12,LCK13,MSDRN20,Kla24,KLK26}.

Our results show that the large-scale loops characteristic of
anomalous looping are not generated passively by ocean currents
\cite{LLL06}, nor do they arise from chirality \cite{BeDiL16} or from
odd diffusion \cite{HEM21}. Instead, our analysis suggests temporally
correlated movement by the turtles within a bounded foraging region,
which corresponds to a novel type of biological activity generalising
a widely known active Brownian particle model
\cite{KMDL14,Sza14}. Environmental conditions, including bathymetry,
sea surface temperature and chlorophyll-a concentration \cite{RSS25},
as well as regional currents, are likely to shape the foraging
boundaries in this region. The oscillatory velocity correlations
provide a statistical signature of the looping dynamics within this
region. In that sense, our discovery of anomalous looping connects the
three fields of ME, AP and AD: The loops are adapted to the
environmental constraints, actively biologically generated and
anomalous in terms of the associated diffusive spreading.

This novel movement mode raises the question of whether loop-based
active movement can enhance foraging in bounded heterogeneous
habitats. Based on our results, we conjecture that anomalous looping
might provide an optimal search strategy to exploit renewable,
patchily distributed moving resources within a spatially constrained
favorable habitat. However, as is well-known from the literature on
L\'evy walks, assessing search efficiency is difficult (see, e.g.,
Ref.~\cite{LTBV20} and the ensuing discussion). Exploring anomalous
looping in view of search efficiency will thus require detailed
theoretical and empirical analysis. Interestingly, similar loop- or
spiral-like movement motifs are known within the context of search and
rescue operations (as expanding square search
\cite{Frost96}). Spiral-like search patterns have in turn famously
been observed for desert ants \cite{Wehn20}, while progressively
larger randomised loops anchored to a home base, termed foray search,
have been reported for butterflies \cite{CZR03}. There are also
suggestions that oscillatory movement may be deeply embedded in neural
control systems of organisms \cite{Cheng22}. These cross-links suggest
that anomalous looping may be associated with a broader class of
loop-based search behaviours. How widespread such dynamics are, and
under which ecological conditions they are favoured, remain important
open questions.

\backmatter
\bmhead{Acknowledgments} R.K. thanks F. Hanke for helpful
discussions. We also thank NGOs in Cabo Verde who
supported the logistics for the tracker deployment.

\section*{Declarations}

\subsection*{Conflict of interest}
The authors declare no competing interests.

\subsection*{Data availability}
All data supporting the findings of this study are available in the manuscript and
its Supplementary Information. Further data can be obtained from the corresponding
author on reasonable request.

\subsection*{Code availability}
Code in support of the finding of this study can be obtained from the
corresponding author on reasonable request.

\subsection*{Author contributions}
Author contributions are defined based on the CRediT (Contributor
Roles Taxonomy) and listed alphabetically.  Conceptualisation: R.K.,
V.V.P. Data curation: P.R.-T. Formal analysis: All.  Funding
acquisition: C.E., V.V.P. Investigation: R.K., V.K., V.V.P. Field
work: C.E. Methodology: R.K., V.K., V.V.P. Project administration:
R.K., V.V.P. Resources: C.E., V.V.P. Software: V.K. Supervision: C.E.,
R.K., V.V.P. Validation: R.K., V.K., V.V.P.  Visualisation:
V.K. Writing – original draft: R.K. Writing – review and editing: All.

\section*{Methods}

\subsection*{Data analysis}
The original data set consisted of ten loggerhead sea turtles (Caretta
caretta) nesting in Cabo Verde. They were tracked using
satellite-relayed telemetry (see SI Sec.1.1 and 1.2 as well as
Ref.~\cite{RSS25} for details of the trackers and turtle
characteristics). All ten turtles show oceanic foraging
behaviour, i.e., staying away from shallow coastal waters during their
foraging phase. This oceanic feeding is most common for
turtles nesting in Cabo Verde.
The dataset is composed of nine females and one male. The trackers
transmitted data for durations comprised between 123 and 1094 days,
with an average of 492 ± 370 days, delivering a total of 9759
locations. The data consist of latitude and longitude coordinates
irregularly spaced in time. We have used linear interpolation to
regularise the time series for the statistical analysis as well as for
the data driven modelling. The sampling interval for the linear
interpolation was computed by taking the average of the time intervals
along each time series. Before performing data analysis, we performed
three stages of data cleaning, see SI Sec.1.3. Full details of the
data analysis yielding the results displayed in Figs.~\ref{fig:Fig2} to
\ref{fig:Fig4} are given in the SI.

\section*{Computer simulations}
Here we outline the numerical procedure of how we solved our
stochastic model, the generalised Langevin equation constructed from
data described in the text, for generating the simulated data shown in
Figs.~\ref{fig:Fig3} and \ref{fig:Fig4}. (1) We generated
uncorrelated samples from the chosen velocity probability
distribution. For sake of simplicity, we chose to ignore the
exponential deviations in the tails and generated purely Gaussian
samples of the velocities. (2) We then fitted the velocity
autocorrelation function to the experimental data as described in the
text and used the result as the input for the covariance matrix. (3)
We used Cholesky decomposition for the covariance matrix to generate
the desired correlated Gaussian velocities. (4) For implementing the
boundary conditions in the longitudinal direction reflecting the
impact of environmental conditions, we used an asymmetric harmonic
potential with different parameters for both branches of the
potential. The coastline we modelled by a steep potential while the deep sea boundary was reproduced by a shallow one. (5) These
parameters of the potentials were adapted by trial and error to match
the experimental results.

\onecolumn
\newpage
\setcounter{section}{0}
\setcounter{figure}{0}
\setcounter{table}{0}
\setcounter{equation}{0}
\renewcommand{\thesection}{\arabic{section}}
\renewcommand{\thefigure}{S\arabic{figure}}
\renewcommand{\thetable}{S\arabic{table}}
\renewcommand{\theequation}{S\arabic{equation}}
\renewcommand{\thesubsection}{\thesection.\arabic{subsection}}
\onecolumn
\newgeometry{a4paper, left=1.5in, right=1.5in, top=0.7in, bottom=0.7in}
\setkeys{Gin}{keepaspectratio, max height=0.83\textheight, max width=\linewidth}

\begin{center}
{\Large\textbf{Supplementary Information:\\Active movement of foraging sea turtles generates anomalous looping}}
\end{center}
\tableofcontents

\section{Statistical data analysis for all the turtles}
\subsection{Introduction}
Roman-Torres et al.~\cite{RSS25} tracked with satellite transmitters 15 adult loggerhead sea turtles (\textit{Caretta caretta}) nesting at the Cabo Verde Archipelago. Basic details of all these turtles are summarised in Table~\ref{tab:turtle_summary} included below. Among them, 12 were females and 3 were males.  Two turtles, namely, Manga and  Nhanha were lost and three turtles (Bolocha, Fra,  and  Ze)  were neretic (i.e. foraging in the shallow water with depths down to around 40 metres). That leaves 10 oceanic turtles that forage in the region of deep bathymetry (with ocean floor depths exceeding 3000m) and feed on a pelagic prey, specifically, epipelagic organisms. The chlorophyll concentration is an indicator of phytoplankton bloom, which serves as a food source for epipelagic predators. Accordingly, it can be considered as a proxy variable for the prey concentration of the turtles. The turtles were tagged during the nesting seasons of 2011, 2012, and 2013 across the following four islands: Sal, Boa Vista, São Vicente, and Fogo.  A turtle has three behavour phases: Nesting, migration and foraging. We focus on the foraging phase.
After spending a nesting period on an island, the sea turtles move towards a feeding area. Once there, the turtles switch to foraging behaviour, displaying a motion with large-scale loops that we denote as "anomalous looping", see Fig.1 in the main text.

\begin{table}[htbp]
\centering
\resizebox{\textwidth}{!}{
\begin{tabular}{|l|l|l|l|r|r|r|}
\hline
\textbf{Turtle Name} & \textbf{Sex} & \textbf{Origin Island} & \textbf{Foraging Strategy} & \textbf{Days Transmitted} & \textbf{Distance (km)} & \textbf{Data Points} \\ \hline
\multicolumn{7}{|c|}{\textbf{Oceanic Foragers}} \\ \hline
Nusco & Female & Boa Vista & Oceanic & 241 & 4,658 & 1,002 \\ \hline
Goody & Female & Boa Vista & Oceanic & 258 & 5,297 & 1,188 \\ \hline
Catchupa & Female & Sal & Oceanic & 147 & 2,842 & 447 \\ \hline
Papaya & Female & Sal & Oceanic & 123 & 2,959 & 422 \\ \hline
Bemvinda & Female & S\~{a}o Vicente & Oceanic & 1,094 & 17,774 & 1,861 \\ \hline
Kika & Female & S\~{a}o Vicente & Oceanic & 850 & 10,580 & 1,491 \\ \hline
Kamoka & Female & Fogo & Oceanic & 922 & 16,749 & 2,595 \\ \hline
Olympia & Female & S\~{a}o Vicente & Oceanic & 154 & 3,454 & 677 \\ \hline
Mokamba & Female & Fogo & Oceanic & 399 & 8,792 & 1,354 \\ \hline
Mingo & Male & Boa Vista & Oceanic & 732 & 13,604 & 1,330 \\ \hline
\multicolumn{7}{|c|}{\textbf{Neritic Foragers}} \\ \hline
Bolacha & Female & Boa Vista & Neritic & 395 & 4,385 & 1,219 \\ \hline
Fra & Male & Boa Vista & Neritic (Resident) & 222 & 1,349 & 530 \\ \hline
Ze & Male & Boa Vista & Neritic (Resident) & 51 & 378 & 90 \\ \hline
\multicolumn{7}{|c|}{\textbf{Lost / Insufficient Data}} \\ \hline
Manga & Female & Sal & Lost & 1 & 12 & 6 \\ \hline
Nhanha & Female & Fogo & Lost & NA & NA & NA \\ \hline
\end{tabular}
}
\caption{An overview of data for 15 tracked loggerhead sea turtles from the Cabo Verde archipelago. Turtles are categorized by foraging strategy (oceanic, neritic, or lost due to transmission failure/short duration). The total dataset consisted of 14,212 locations.}
\label{tab:turtle_summary}
\end{table}

Based on statistical data analysis, we construct a stochastic model that reproduces the foraging dynamics of the  sea turtles.

\subsection{Data collection}
\begin{figure}   
    \centering
 \makebox[\textwidth][c]{ \includegraphics[width=1.0\linewidth]{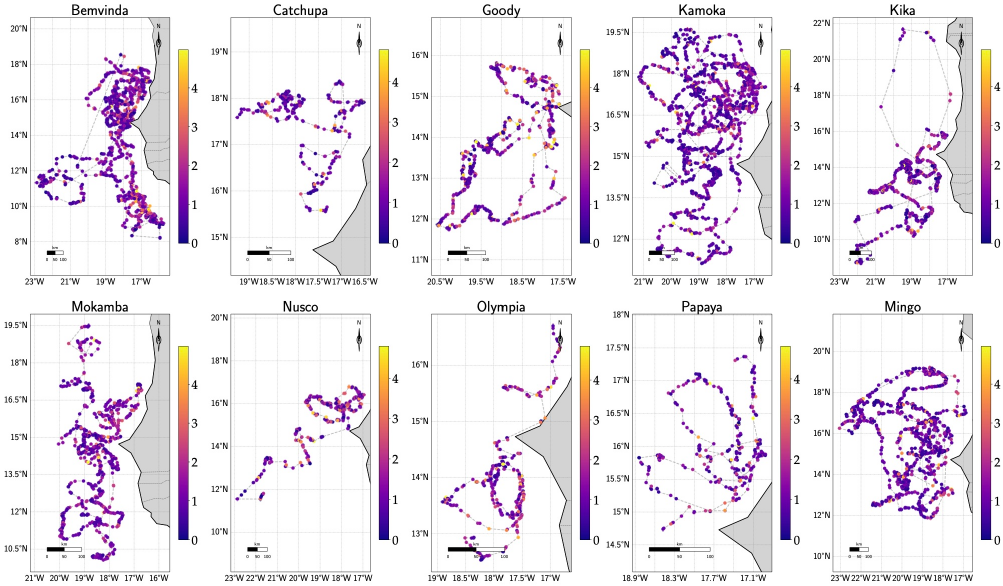} }
    \caption{Trajectories in the foraging phase for the 10 loggerhead sea turtles identified before. One can see that all turtle paths display loops. No preferred area for the looping can be identified. Some turtles even take longer excursions away from the coastline by generating loops. The color code shows the magnitude of the speed $s$ (measured in km/h) at each point.}
    \label{fig:FigS1}
\end{figure}

The Argos Satellite Tracking System has been used to collect the movement data of the loggerhead sea turtles tagged with GPS relayed tags recording the latitude and the longitude coordinates. Since the signal could only be recorded when the turtle was near the surface, the obtained time series data for latitude and longitude are not regular. Figure~\ref{fig:FigS1} shows the trajectories of 10 of the turtles during the foraging phase.

\subsection{Data cleaning}
\begin{figure}[H]
    \centering
     \includegraphics[width=1.0\linewidth]{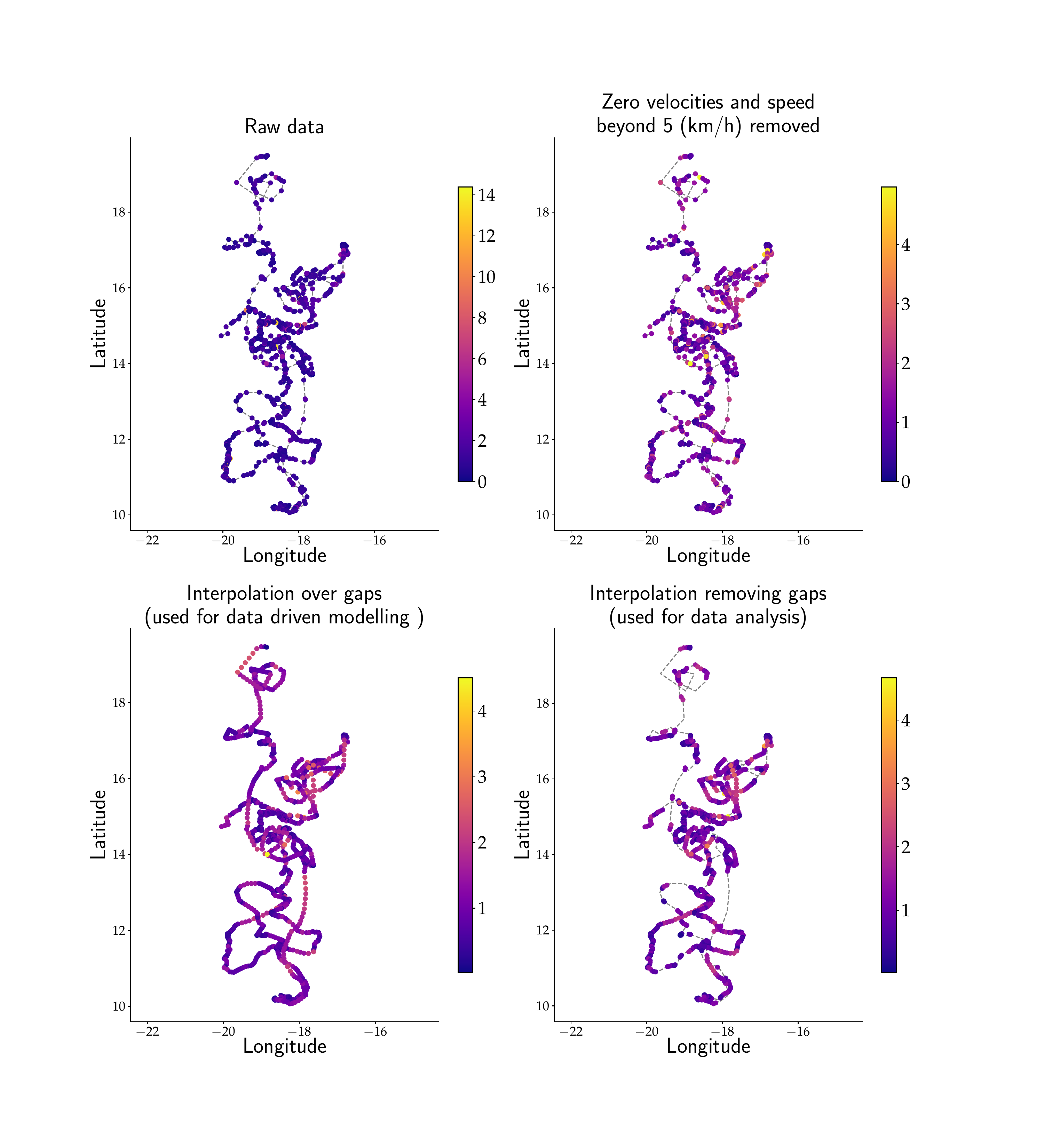} 
    \caption{Path of the turtle Mokamba, colour coded by speed $s$ (measured in km/h). Top left: trajectory plotted from the raw data.  Top right: We remove extreme velocities and duplicates of zero velocities. Bottom left: Interpolation over the gaps is done (the extreme velocities are also removed). Bottom right: Interpolation from irregular to regular time series after removing the gaps.}
    \label{fig:FigS2}
\end{figure}
\begin{figure}[H]
    \centering
    \makebox[\textwidth][c]{
          \includegraphics[width=1.2\linewidth]{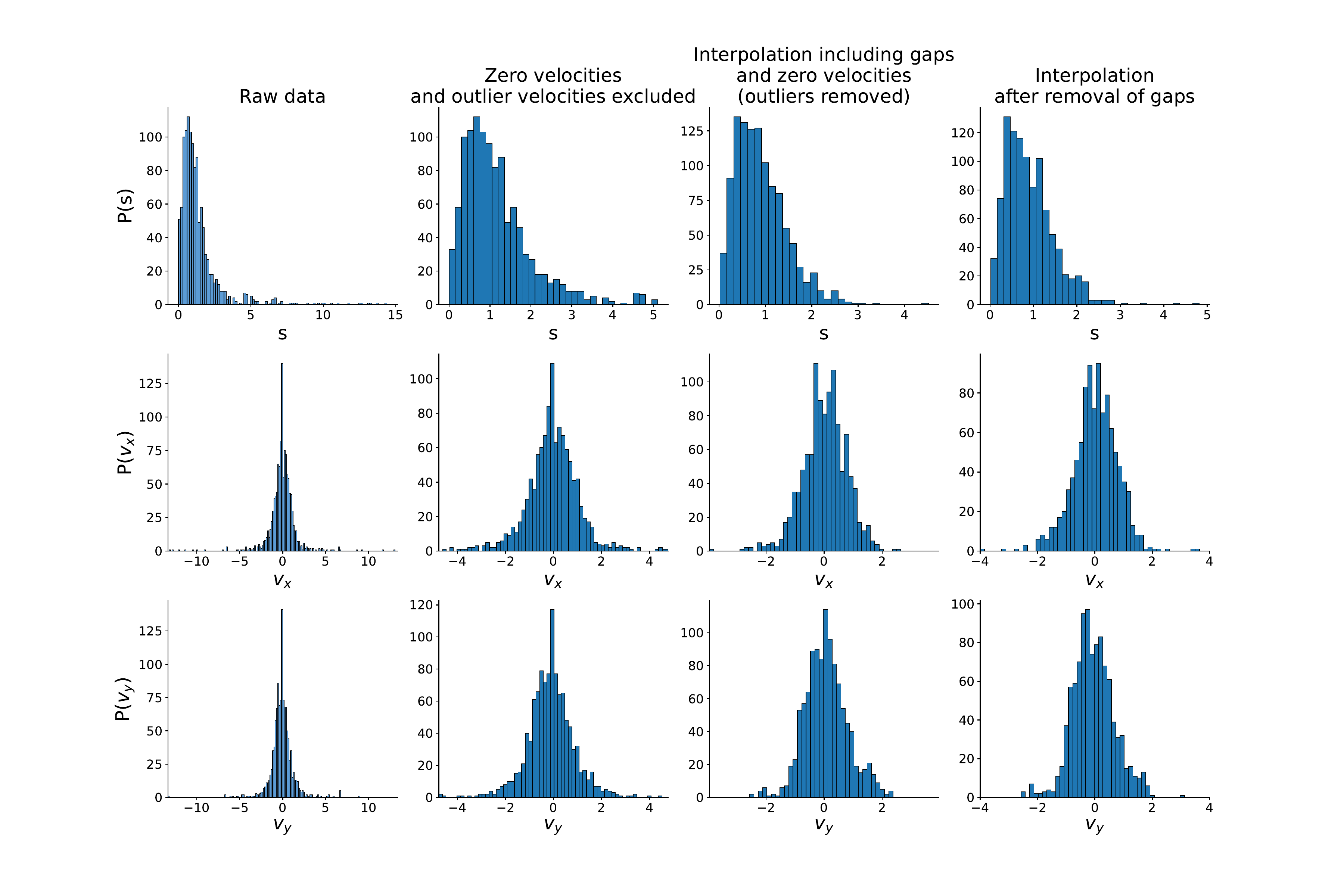} 
    }
    \caption{There are three stages of the data cleaning. The first stage involves removing any duplicates,  zero velocities and extreme velocities.
    From the first stage cleaned data, we independently  perform the 2nd stage and third stage data cleaning and transformation. The first stage cleaned data still consist of irregular time series of the velocities and speeds, which needs to be regularized by interpolation. In the second stage, we interpolate the complete trajectory obtained from first stage cleaning.  However, at times there are large gaps in the time series. Hence, in the third stage we remove these gaps identified via box plot analysis and then split the trajectory into several segments. Within these short segments we interpolate the regular time series from the available irregular series. For the statistical data analysis we have used the third stage. We have also done a similar analysis for second stage and find that it does not produce qualitative differences. We used second stage cleaned data for the data-driven modelling. The speed $s$ and the velocity components ($v_x$, $v_y$) are measured in km/h.}
    \label{fig:FigS3}
\end{figure}
\begin{figure}[H]
    \centering
    \makebox[\textwidth][c]{
         \includegraphics[width=1\linewidth]{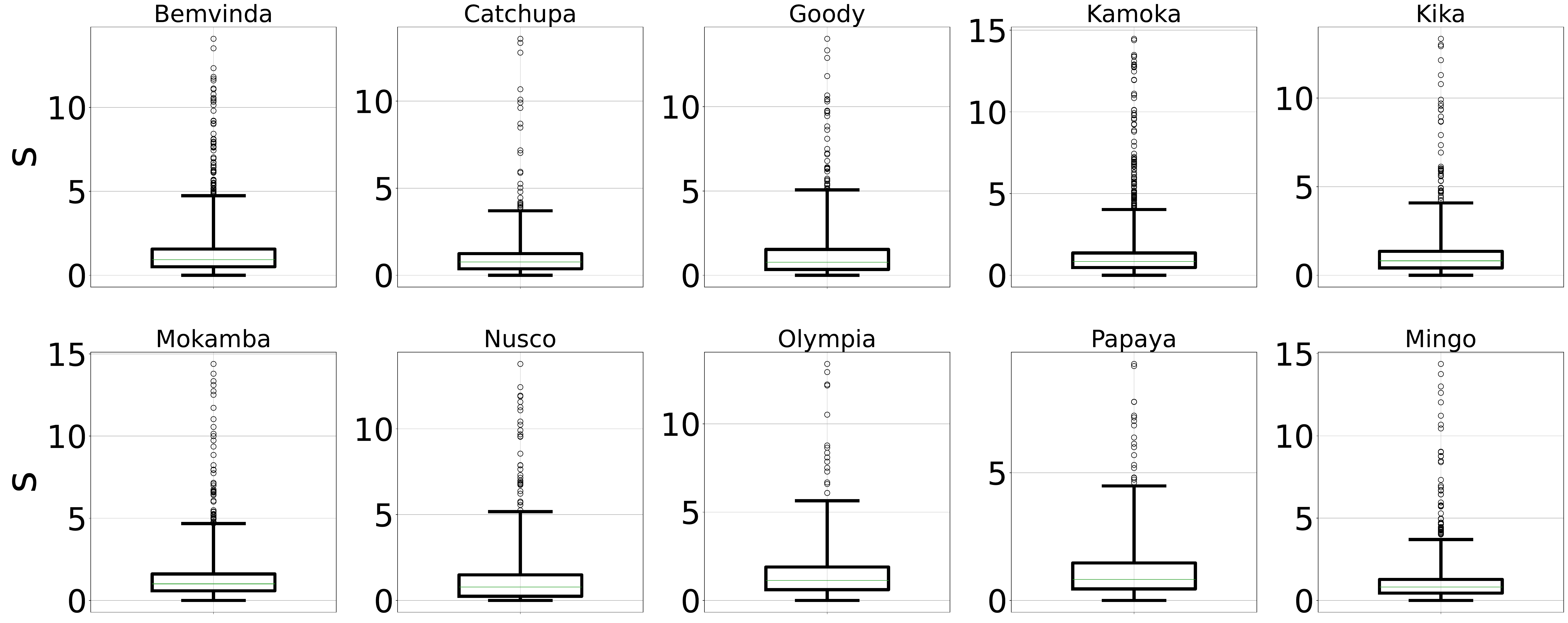} 
    }
    \caption{Box plot analysis of speed data for 10 sea oceanic turtles in the foraging phase. In each subplot in the figure, the central horizontal line represents the median speed, and the box spans the interquartile range (IQR) from the 25th to the 75th percentile. The whiskers extend to a distance of $3.0 \times \text{IQR}$ (whisker 3.0), capturing the bulk of the data within the IQR range while identifying only extreme values as outliers. Points beyond these whiskers are shown as individual circles, highlighting unrealistically high speeds.}
    \label{fig:FigS4}
\end{figure}
\begin{figure}[H]
    \centering
    \makebox[\textwidth][c]{
         \includegraphics[width=1\linewidth]{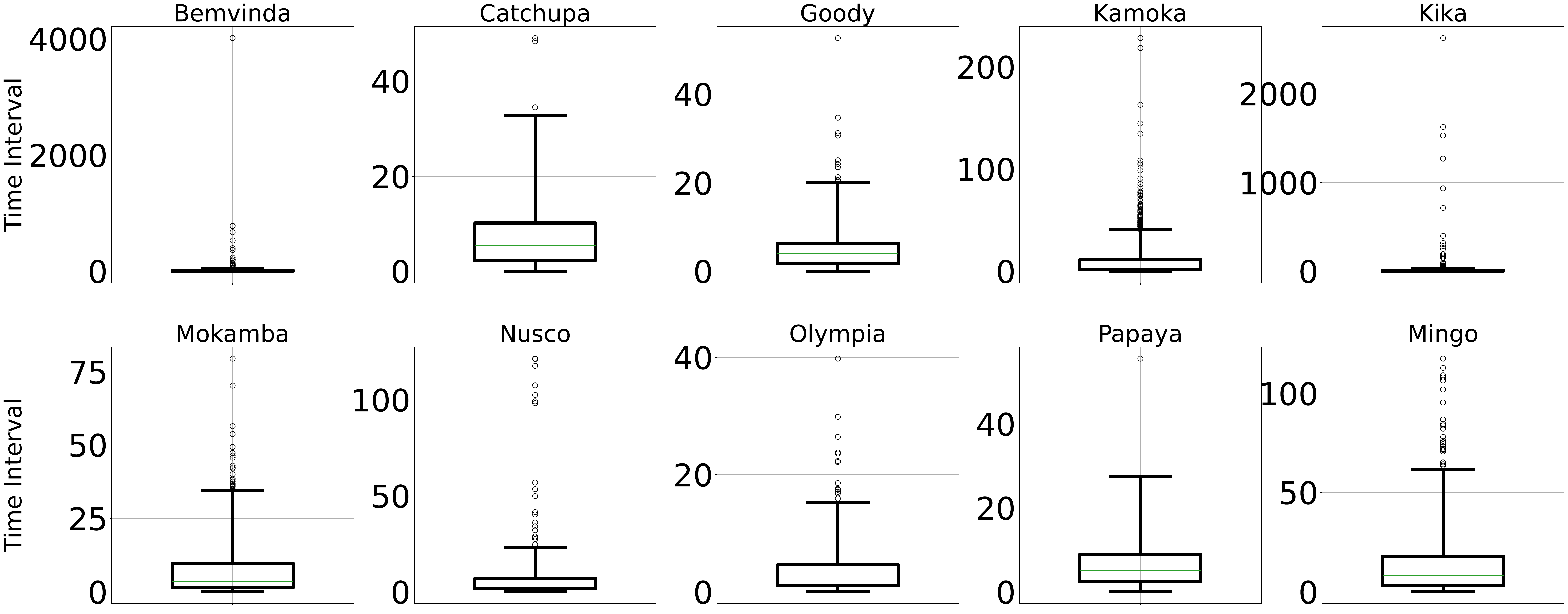} 
    }
    \caption{Box plot analysis of the time intervals between two consecutive data points. The time series of longitude and latitude coordinates is irregular since the time interval between two consecutive data points is irregular.
    One can see the presence of big gaps, especially for Bemvinda and Kika.}
    \label{fig:FigS5}
\end{figure}

The turtle data sets exhibit three types of artefacts: Spurious zero velocities values, outlier gaps (between two consecutive co-ordinates) and high velocities. These anomalies are removed before the data are analysed. Corresponding paths are shown for the example of the turtle Mokamba in Fig.~\ref{fig:FigS2}.
First, sometimes positions were obtained within time spans of seconds, where the turtle barely moved, indicating a multiple recording by the GPS receiver. We thus removed multiples of zero velocities recorded for less than 5 minutes, as otherwise these values would significantly distort the corresponding velocity distributions. This is illustrated in the histogram distributions of the speed and the velocities ($v_x$, $v_y$) in Fig. \ref{fig:FigS3}. Second, the box plot analysis for the speed presented in Fig.~\ref{fig:FigS4} shows that for each turtle there is a percentage of unusually high speeds, which are also not considered for our data analysis, also given that this data is sparse. Third, since longitude and latitude coordinates can only be measured when the turtles swim on the surface, large gaps between two consecutive data points appear. These gaps may also be due to transmission problems between sensor and satellite. Such problems make the time series generically irregular.
By removing the largest gaps identified in the box plot analysis of Fig.~\ref{fig:FigS5}, we split the single trajectory of each turtle into corresponding segments and then linearly interpolate in each segment over the remaining smaller gaps between adjacent points. We then resampled the linearly interpolated data by a time interval defined as
the mean of the time intervals between consecutive data points in the respective segments (third stage data cleaning). Similarly, we have obtained the time interval to be used to interpolate for the second stage data cleaning by taking the mean of all the time intervals between consecutive data points of the trajectory obtained after first stage data cleaning.
After this data cleaning, we computed the probability distribution functions of speed $s$, longitudinal and latitudinal velocities $v_x$, $v_y$ and turning angles $\theta$, as well as the mean squared displacement (MSD), velocity autocorrelation functions (VACF) and velocity cross-correlations among the $x$ and $y$ direction.

\subsection{Analysis of the speed data}
\begin{figure}[H]
    \centering
     \includegraphics[width=1.0\columnwidth]{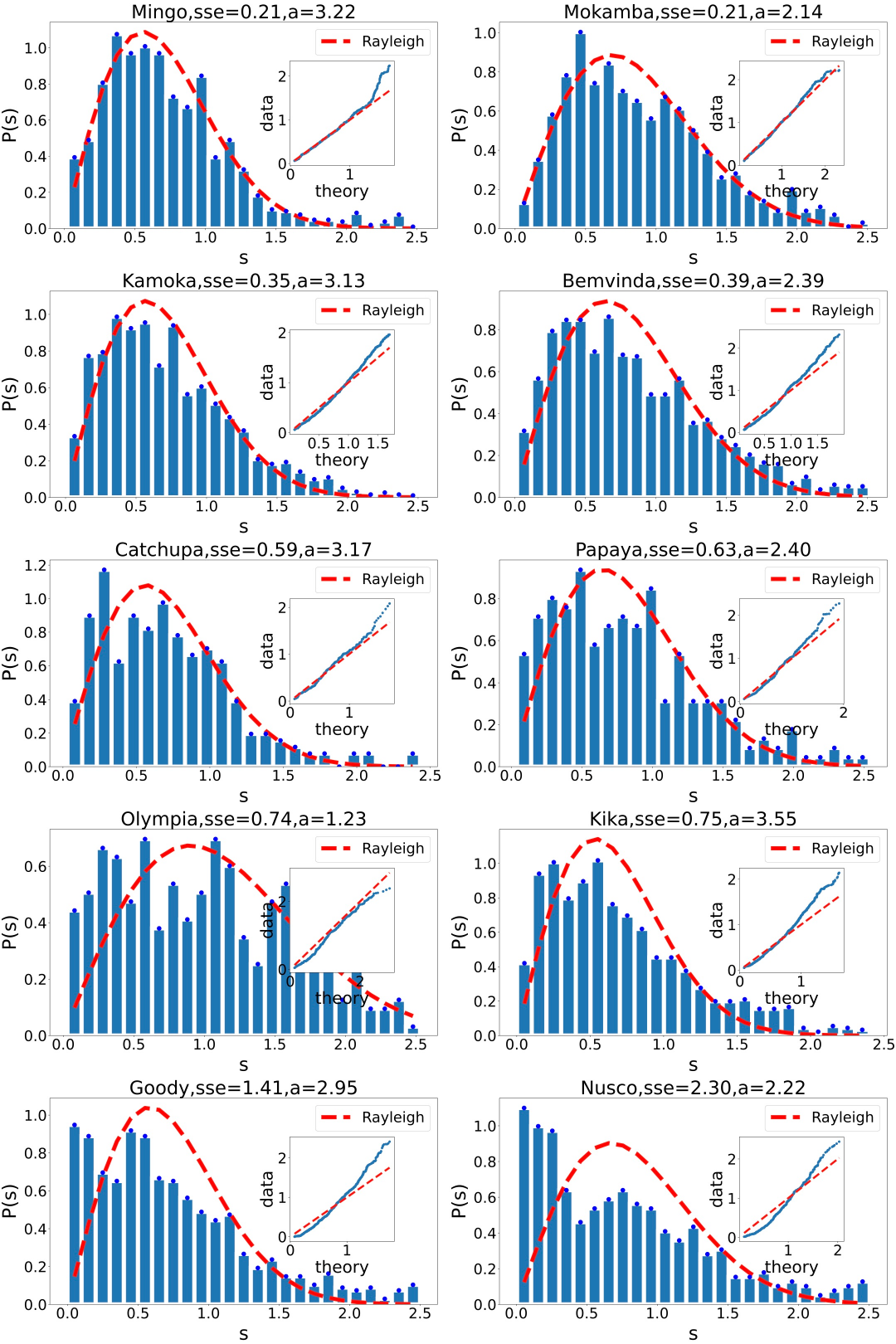} 

    \caption{Rayleigh fit for the speed data. The speed for the turtle is well fitted by Rayleigh distributions, with deviations observed at the tail for most of the turtles, as evident from the Q-Q plots.  The symbol $s$ denotes the speed measured in km/h. The range of the data set is 0-2.5 km/h and the bin size is 0.1 km/h.}
    \label{fig:FigS6}
\end{figure}

\begin{figure}[H]
    \centering
     \includegraphics[width=\columnwidth]{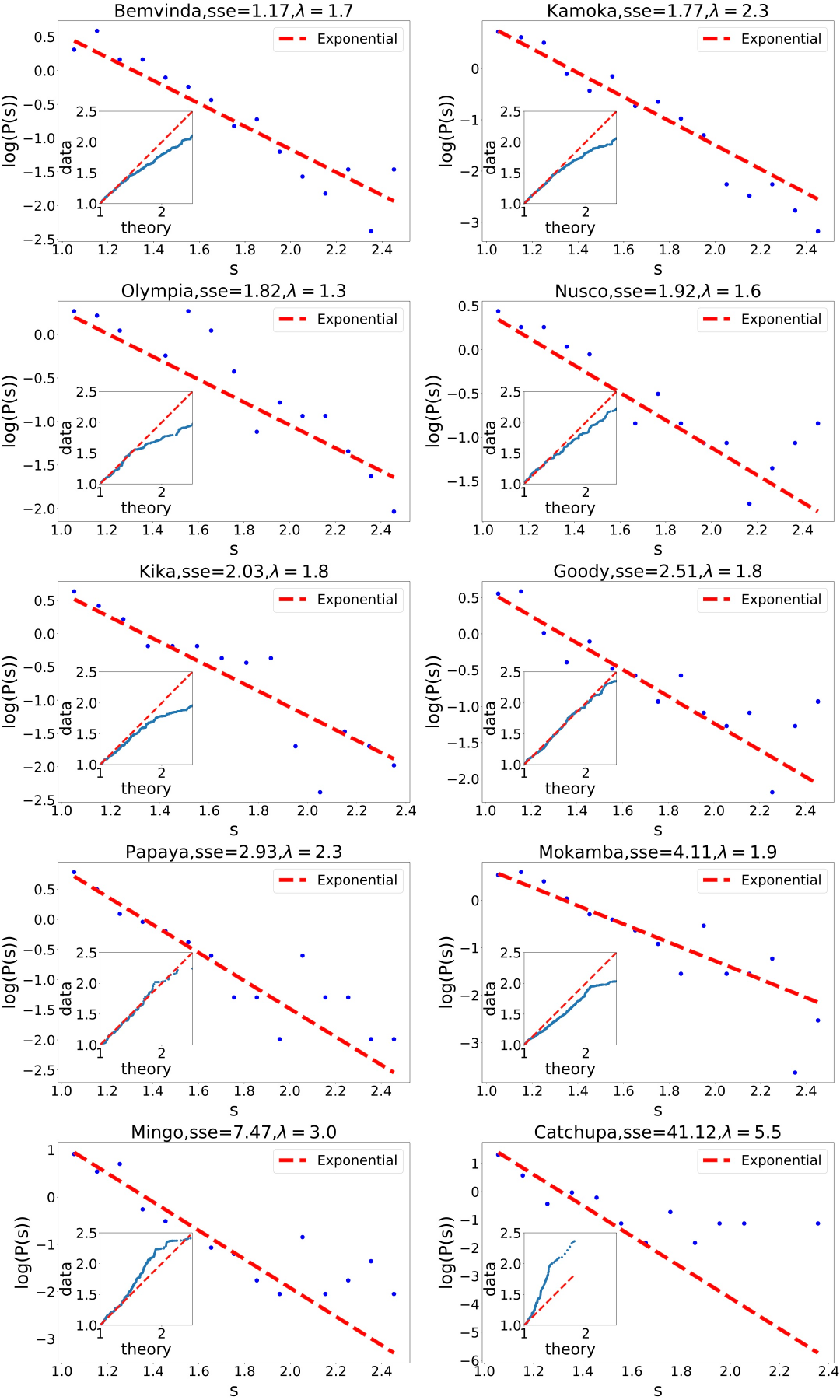} 
    \caption{Exponential fit. For most of the turtles the tail is fitted well by the exponential function, as is clear from the Q-Q insets. The bin size is 0.1 km/h and the range of the data is 1-2.5 km/h. }
    \label{fig:FigS7}
\end{figure}

\begin{figure}[H]
    \centering
     \includegraphics[width=\columnwidth]{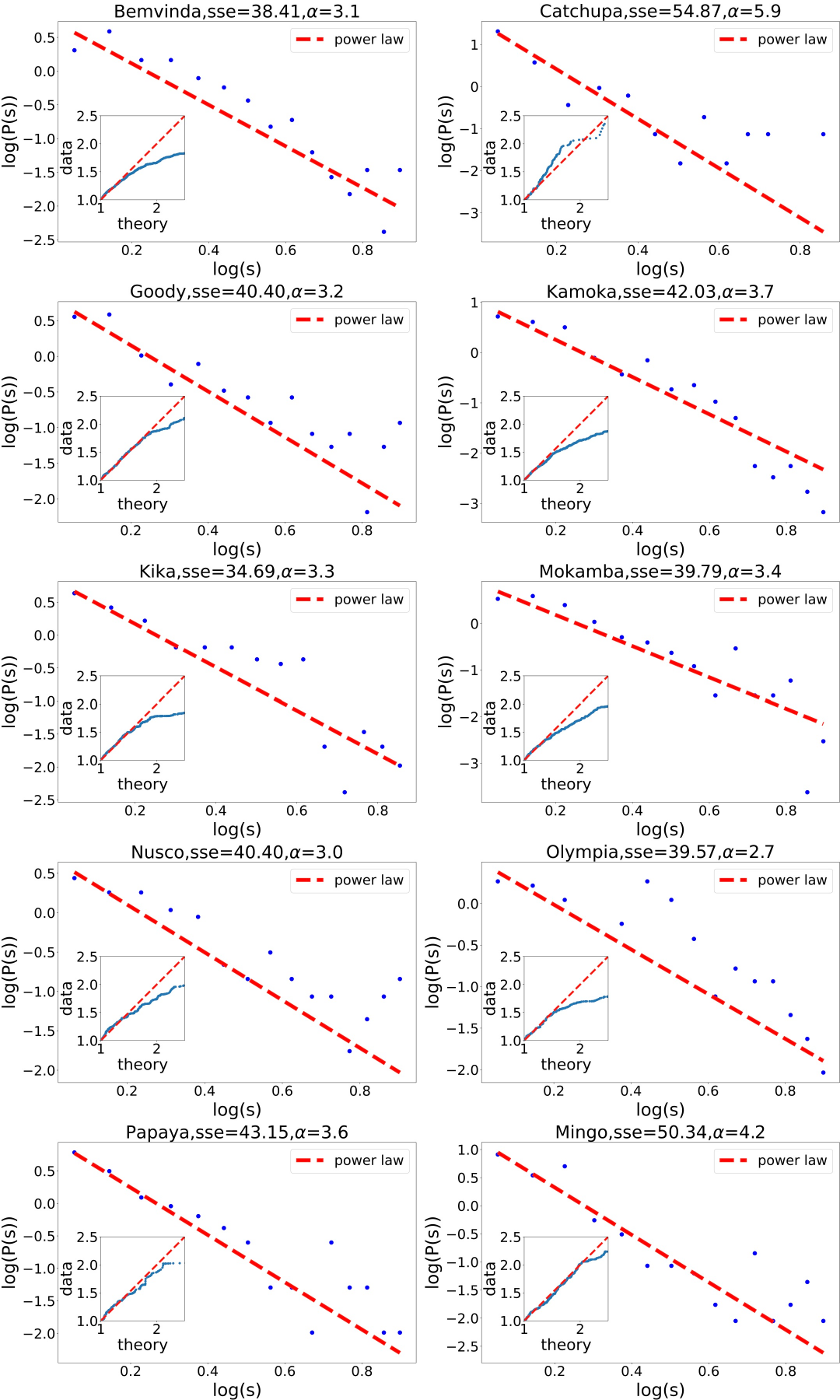} 
    \caption{Power law fit. Compared to the exponential fit, the power law fit shows larger deviations as can be seen from the Q-Q plots. Mingo is an exception, as it takes a long excursion to an island (this excursion rather involves travelling than foraging).}
    \label{fig:FigS8}
\end{figure}

We fitted the speed data with the Rayleigh distribution (two-dimensional Maxwell-Boltzmann distribution), exponential and power-law functions. The experimental data were fitted using a non-linear least squares optimisation routine (\texttt{scipy.optimize.curve\_fit}) in Python.  This function determines the optimal parameters by iteratively minimising the sum of the squared residuals between the experimental data and the fits.
The specific functional forms used for the fits read:
\begin{itemize}
    \item \textbf{Rayleigh:} $f(x) = ax\exp(-a x^2/2)$, where $a$ is a fitting parameter related to the scale.
    \item \textbf{Exponential:} $f(x) = a\exp(-a|x-x_{\text{min}}|)$, fitted to the tail of the distribution beyond a minimum cutoff $x_{\text{min}}$.
    \item \textbf{Power-law:} $f(x) = a|x|^{-b}$, fitted to the same part of the distribution tail.
\end{itemize}
To determine the goodness of the fit, we first visually analysed standard quantile-quantile (Q-Q) plots which are displayed as insets in the following figures.
We find that the Rayleigh distribution is a good fit for the centre of the speed distribution data and even beyond (Figure \ref{fig:FigS6}), with deviations observed primarily at the tail which is particularly evident from the Q-Q plots. Note that the subplots in Figure \ref{fig:FigS6} are arranged in an increasing order of the sum of squared errors (SSE) for each turtle. We have further tested the tail with the exponential (Figure \ref{fig:FigS7}) and the power-law functions. It turns out that the exponential function is a better fit than the power-law (Figure \ref{fig:FigS8}).
The statistical uncertainty (standard error) associated with each fitted parameter is directly estimated from the diagonal elements of the parameter covariance matrix (\texttt{pcov}) returned by the optimisation algorithm.
\subsubsection*{Rayleigh fit for speed data}

\subsubsection*{Exponential fit for speed data}

\subsubsection*{Power law fit for speed data}

\subsection{Analysis of the velocities}
\begin{figure}[H]
    \centering
     \includegraphics[width=\columnwidth]{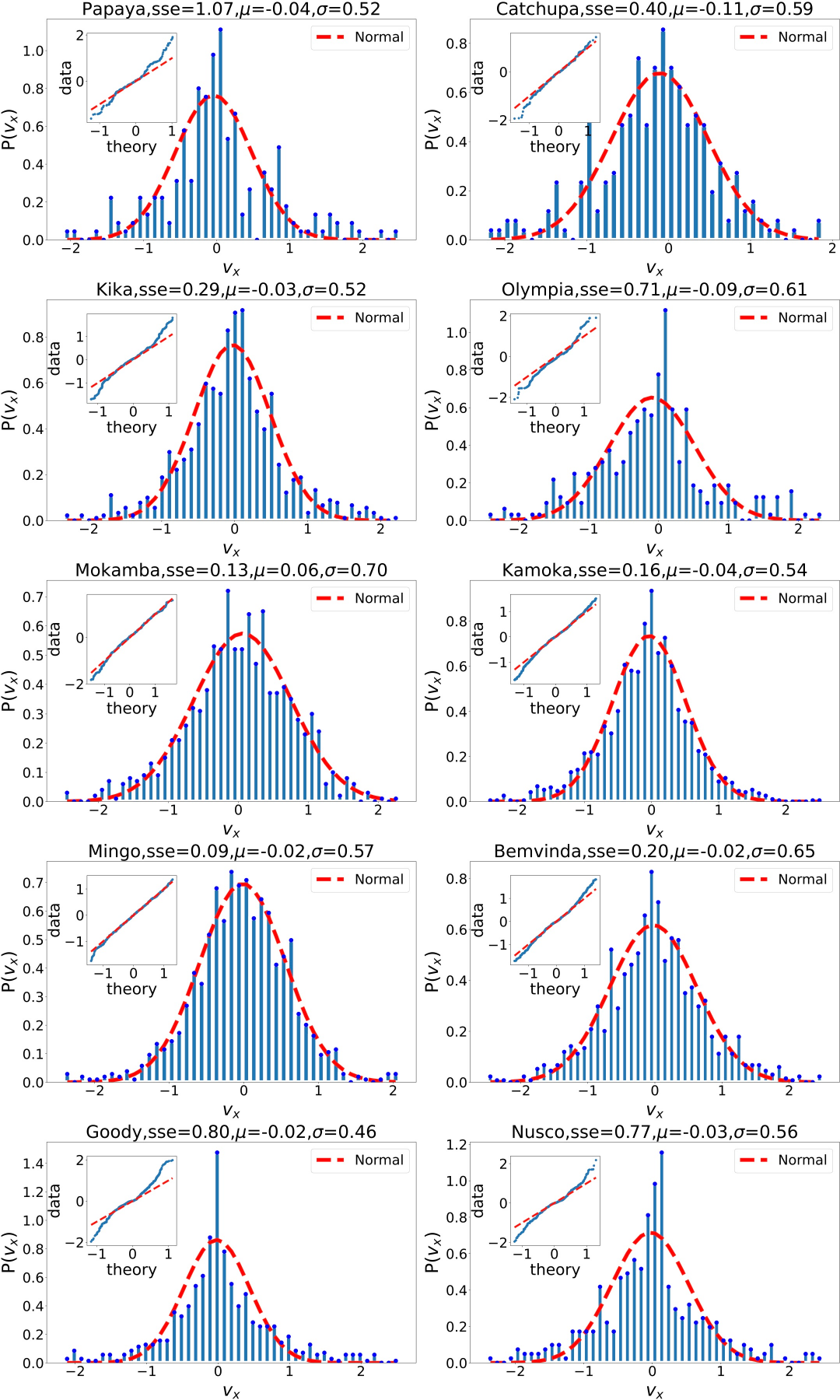} 
    \caption{Normal fit for $v_x$ (measured in km/h). We find the velocity distribution along $x$ direction follows Gaussian at small velocities. This is consistent with the our observation that speed data are well fitted by Rayleigh distribution.}
    \label{fig:FigS9}
\end{figure}

\begin{figure}[H]
    \centering
     \includegraphics[width=\columnwidth]{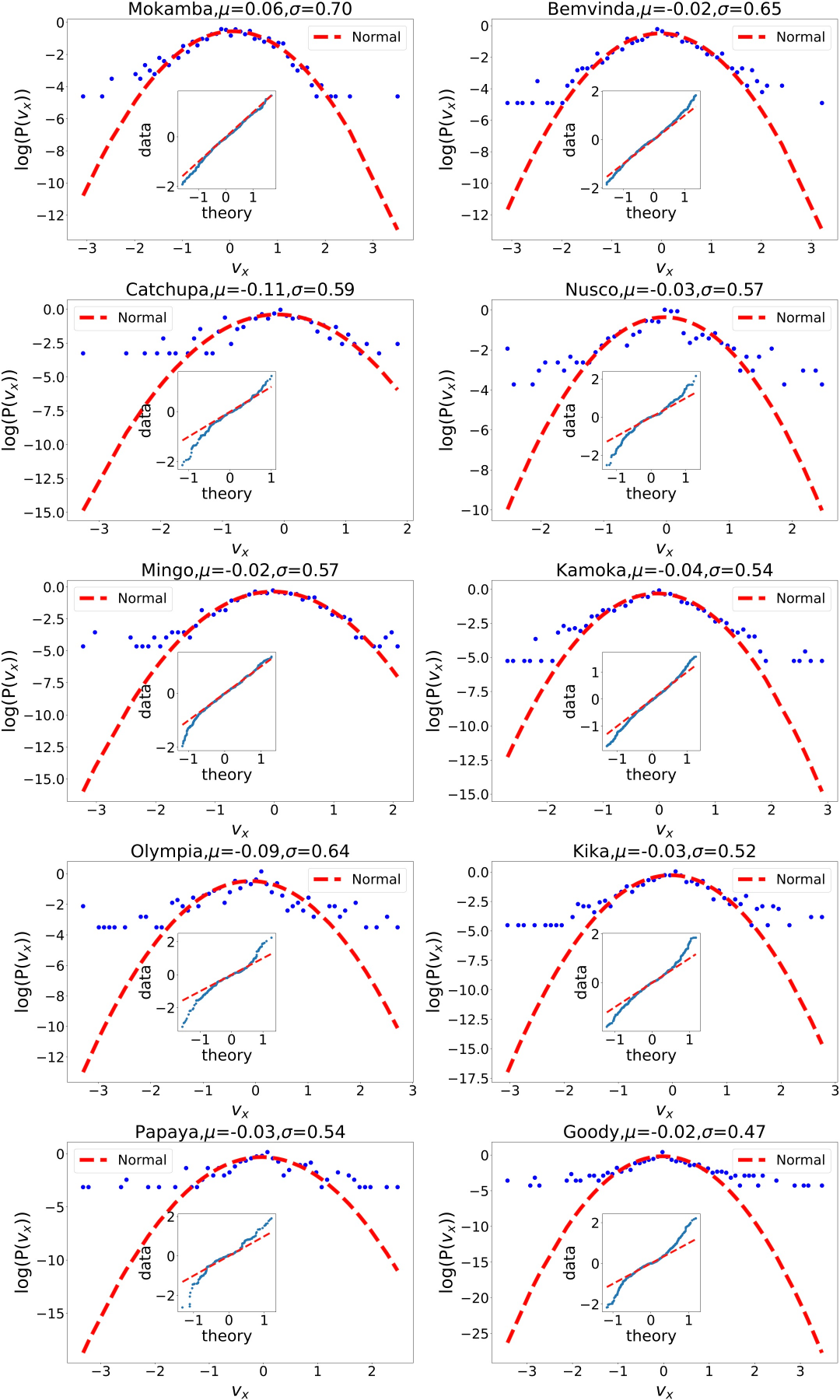} 
    \caption{Log-linear plot for $v_x$ (measured in km/h). The log-linear plots clearly show the deviations from the Gaussian fit (the dashed red lines) at the tails.}
    \label{fig:FigS10}
\end{figure}

\begin{figure}[H]
    \centering
     \includegraphics[width=\columnwidth]{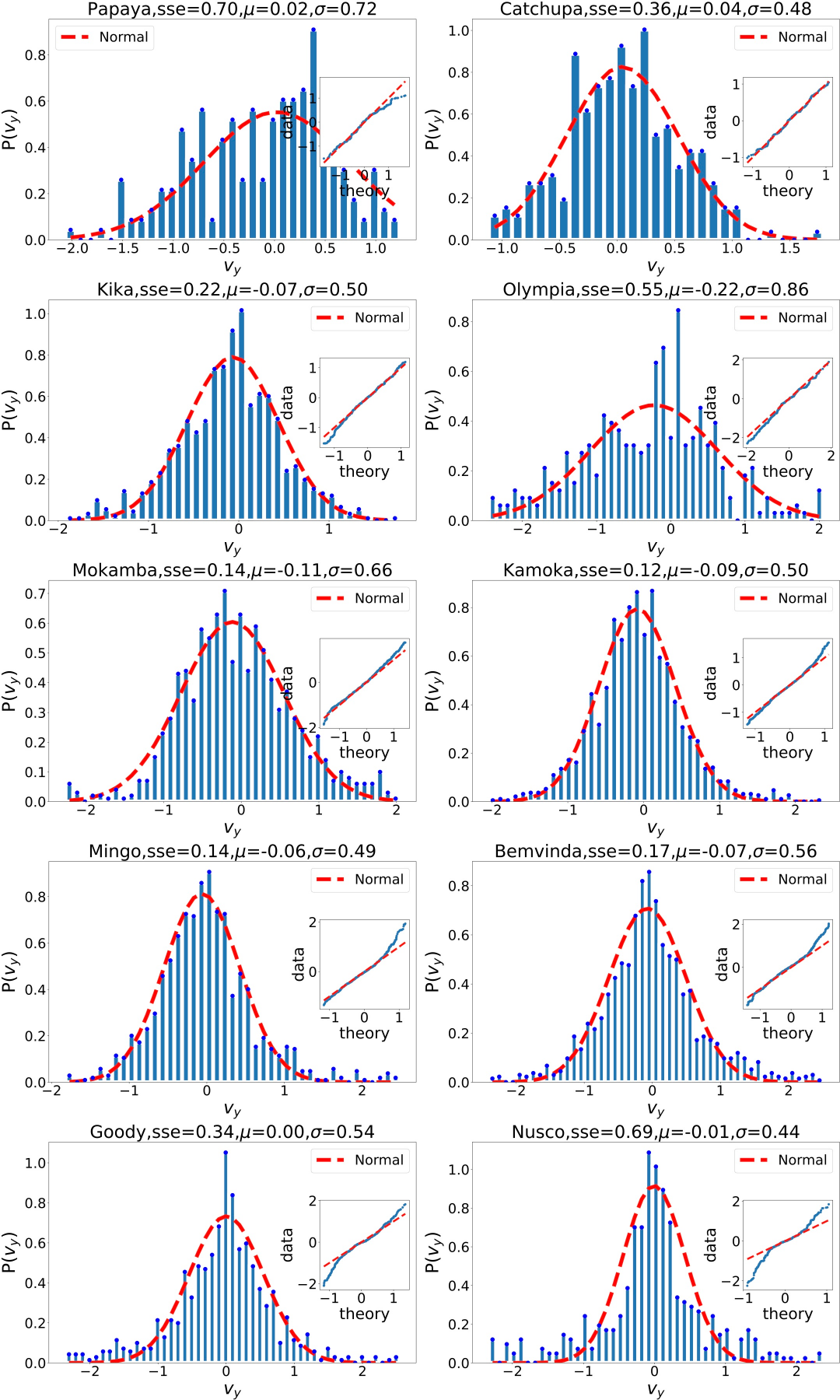} 
    \caption{Normal fit for $v_y$ (measured in km/h). We also find that velocities along $y$ direction are well fitted by Gaussian at the centre. There are longer gaps for some turtles which manifests itself in worse quality fits.}
    \label{fig:FigS11}
\end{figure}

\begin{figure}[H]
    \centering
     \includegraphics[width=\columnwidth]{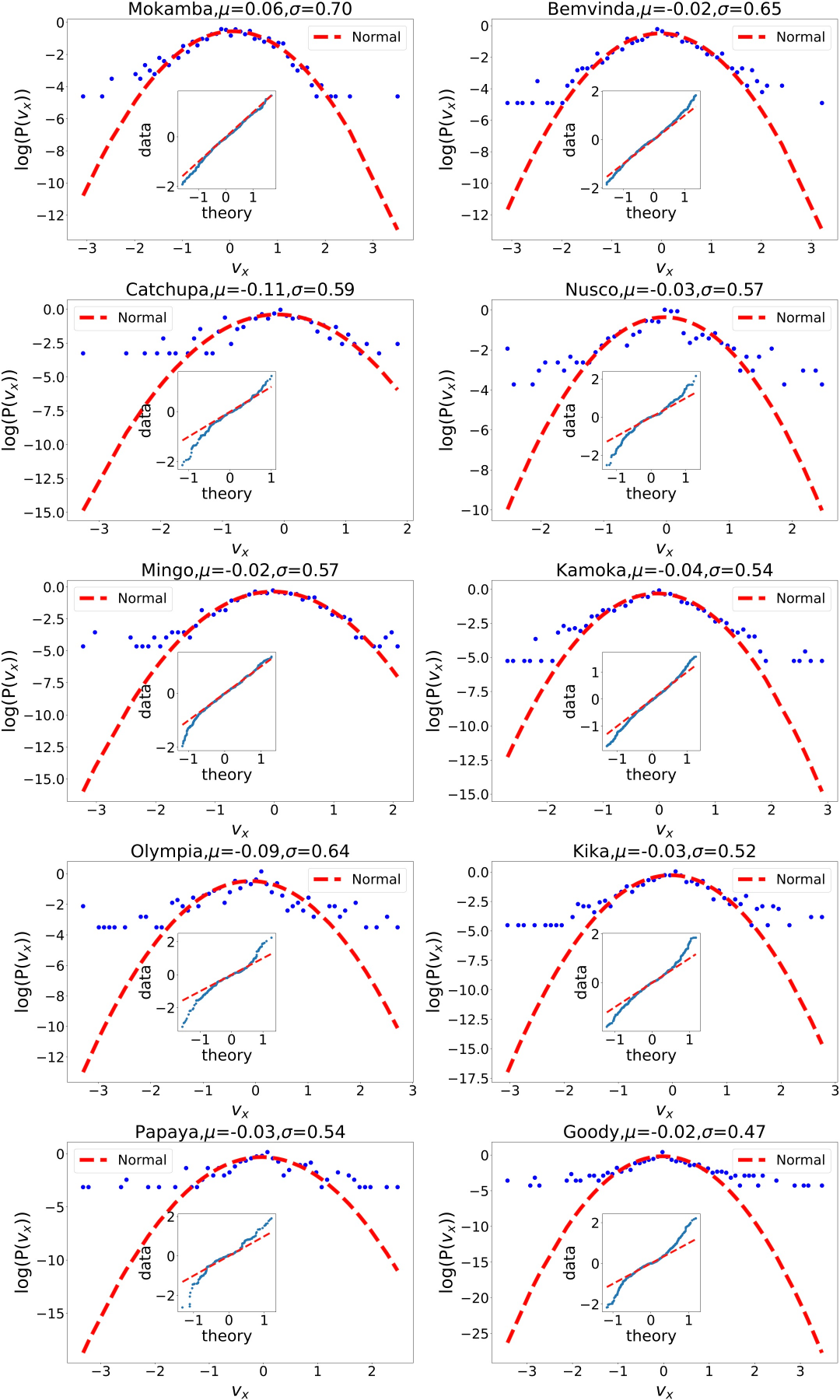} 
    \caption{Log-linear plot for $v_y$ (measured in km/h) with a Gaussian fit shown with dashed red lines.}
    \label{fig:FigS12}
\end{figure}

Corresponding to our result that the tails of the speed distribution are well fitted by an exponential while the centre is well represented by a Rayleigh distribution, we fit the associated velocities distributions in the centre with a Gaussian (normal) distribution,
\begin{equation*}
    f(x) = \frac{1}{\sqrt{2\pi\sigma^2}}\exp\left(-\frac{(x-\mu)^2}{2\sigma^2}\right).
\end{equation*}
The mean $\mu$ and the standard deviation $\sigma$ were deduced via a non-linear least squares optimization routine
(\texttt{scipy.optimize.curve\_fit}). The parameter uncertainties were similarly extracted from the covariance matrix (\texttt{pcov}), with goodness-of-fit measured via the sum of squared errors (SSE) and Pearson correlation coefficient $r$.
The normal distribution indeed fits the centre of the experimental data well (Fig. \ref{fig:FigS9}, Fig. \ref{fig:FigS11}). We show the deviation of the tail behaviour from the normal distribution in log-linear plots (Fig. \ref{fig:FigS10}, Fig. \ref{fig:FigS12}).
The specific symmetric range where the Gaussian well represents the data varies by turtle (for a turtle such as Mokamba, the central Gaussian can extend up to an absolute velocity of 2 km/h). For some turtles there are more significant deviations from good fits, due to the lack of good quality statistics or, in the case of Mingo, due to specific features in the turtle movement. Mingo is the only male in our analysed data, and as explained above, he makes specific excursions to island(s), which is visibly reflected in the profile of the distributions.

\subsubsection*{Normal fit for $v_{x}$.}

\subsubsection*{Normal fit for $v_{y}$. }

\subsection{Statistical data analysis for the second stage cleaned data}
We have also performed data analysis for the second stage cleaned data. For this dataset, we do not remove the gaps and instead perform linear interpolation over the gaps.
We find no significant difference from the third stage cleaned data, specifically, our observation that the tail of the speed data follows an exponential distribution and at small speeds the distribution is fit by a Gaussian.
\begin{figure}[H]
    \centering
     \includegraphics[width=\linewidth]{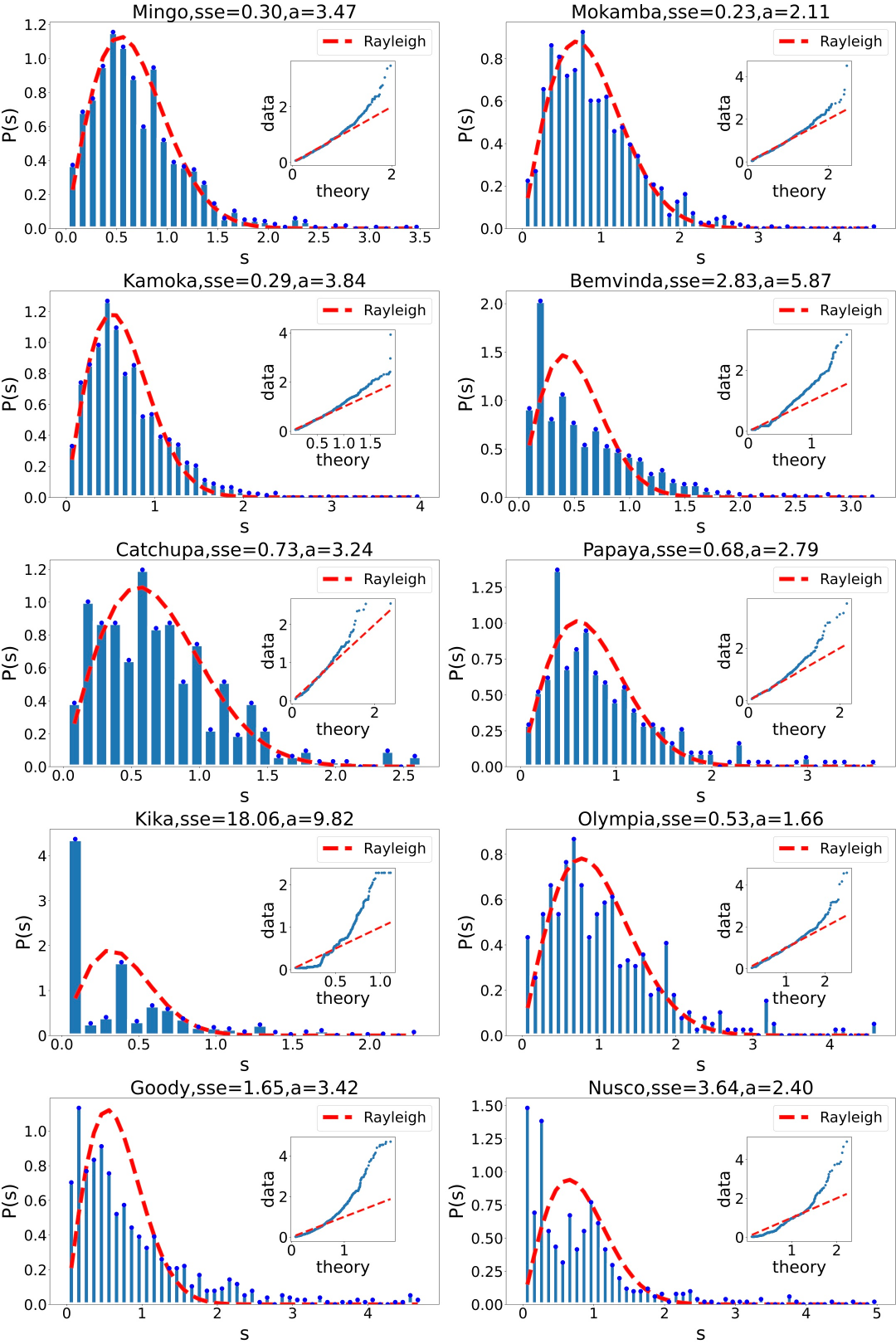} 
    \caption{Rayleigh fit for second stage cleaned data of speed. As we can see from the Q-Q plots, there are deviations at the tail. Hence, we compare the fittings of the tail with the exponential (Fig. \ref{fig:FigS14}) and the power-law function (Fig. \ref{fig:FigS15}).}
    \label{fig:FigS13}
\end{figure}

\begin{figure}[H]
    \centering
     \includegraphics[width=\linewidth]{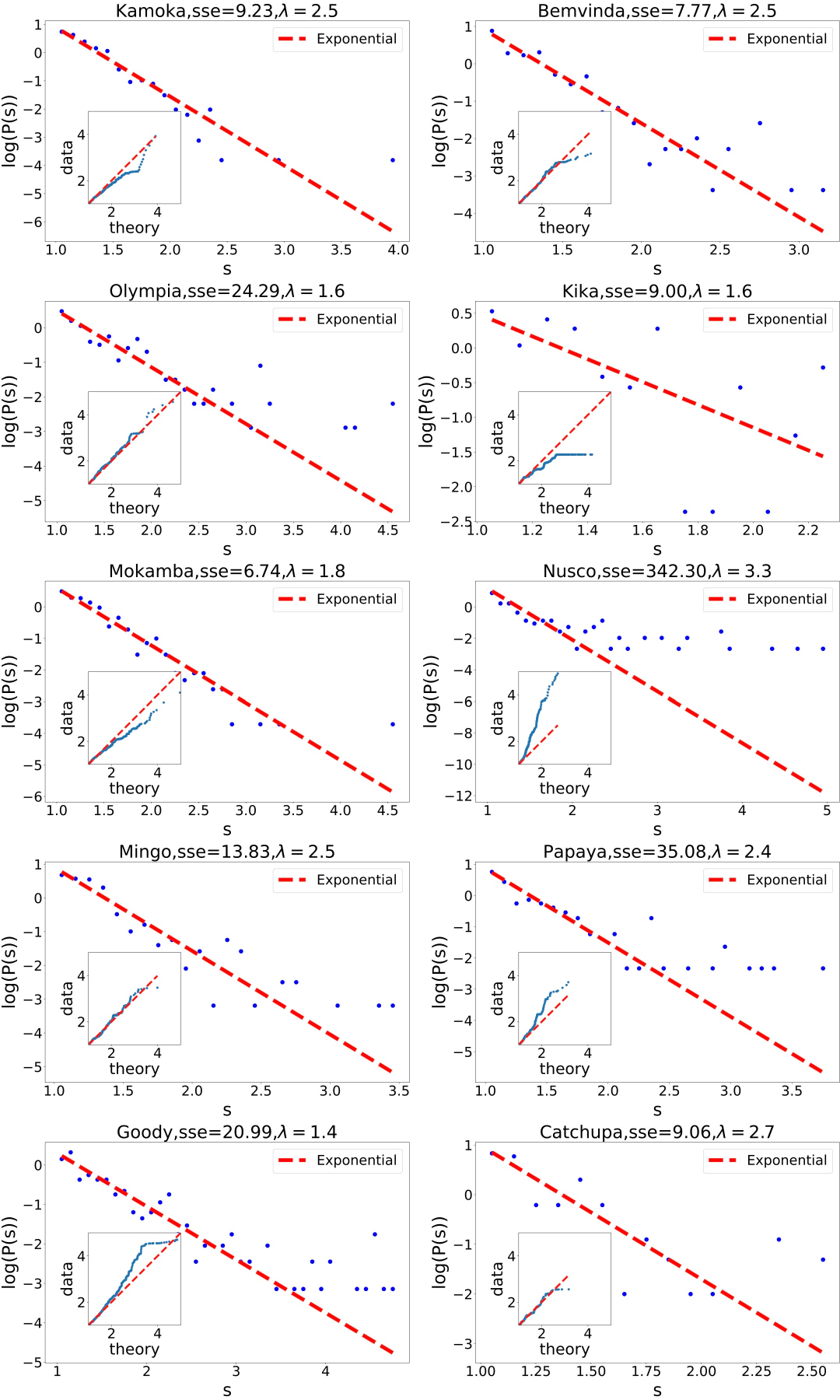} 
    \caption{Exponential fit for the tail of the second stage cleaned data for the speed.}
    \label{fig:FigS14}
\end{figure}

\begin{figure}[htbp]
    \centering
     \includegraphics[width=\linewidth]{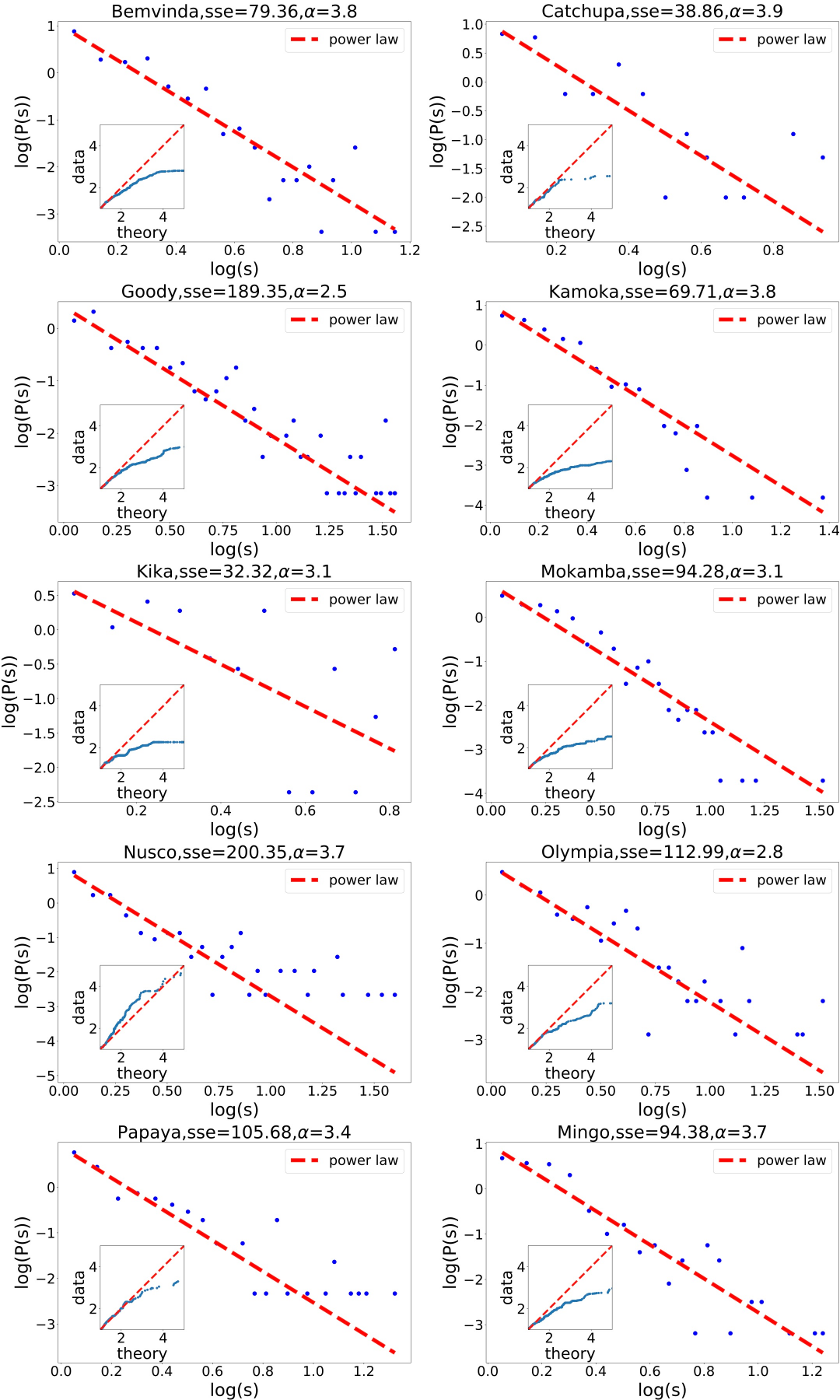} 
    \caption{Power-law fit for the tail of the second stage cleaned data for the speed.}.
    \label{fig:FigS15}
\end{figure}
\subsection{Analysis of the time averaged mean-squared displacement}
\begin{figure}[htbp]
    \centering  \includegraphics[width=\columnwidth]{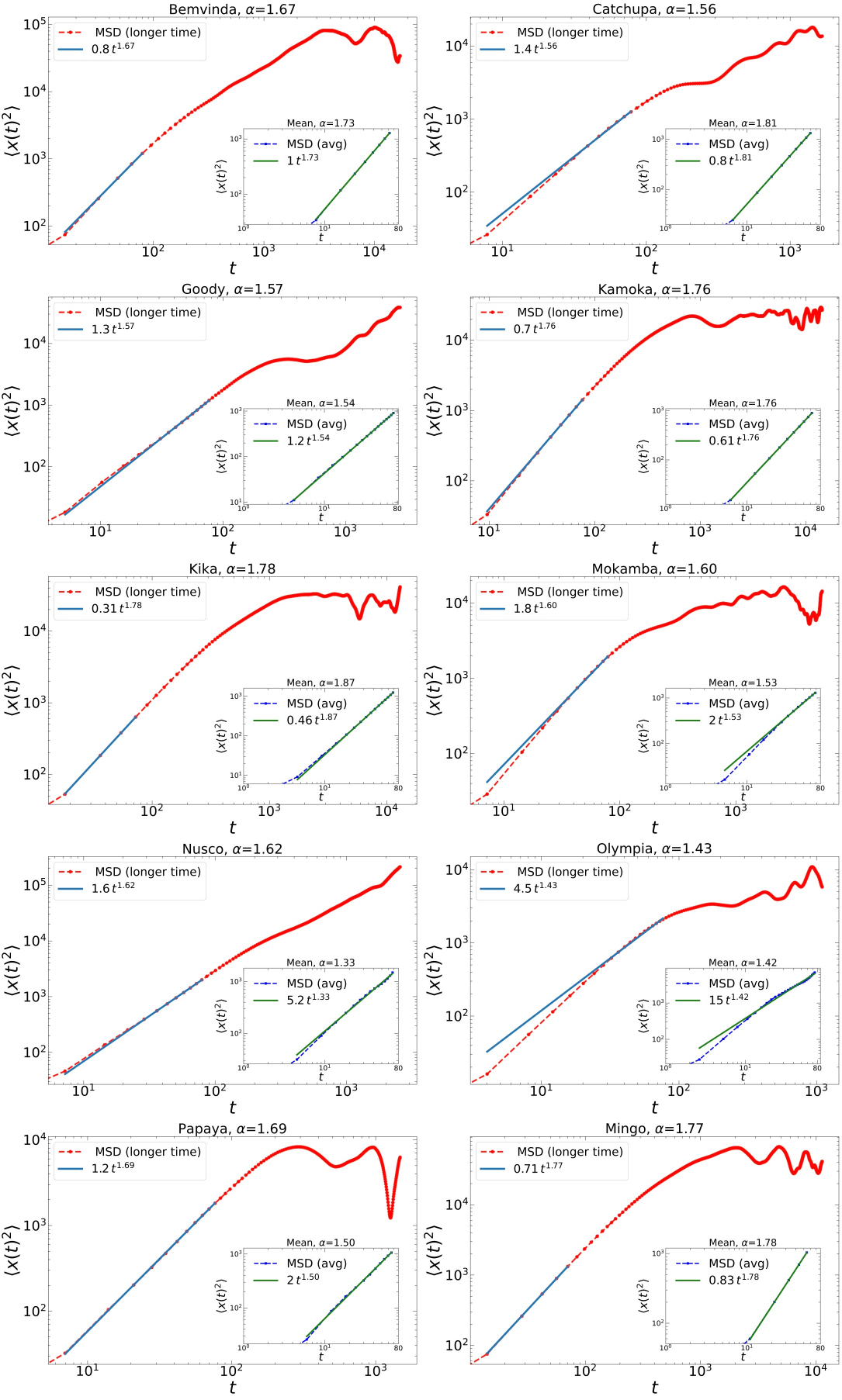} 
    \caption{TAMSD for $\langle x^2(t) \rangle$. As we can see, the exponent $\alpha$ lies within the range $1<\alpha<2$, hence the motion of the turtle in the $x$ direction is superdiffusive. The insets depict the TAMSD averaged over the ensemble of the splitted segments after the removal of the largest gaps.  The displacement $x$ is measured in km.}
    \label{fig:FigS16}
\end{figure}

\begin{figure}[htbp]
    \centering
     \includegraphics[width=\columnwidth]{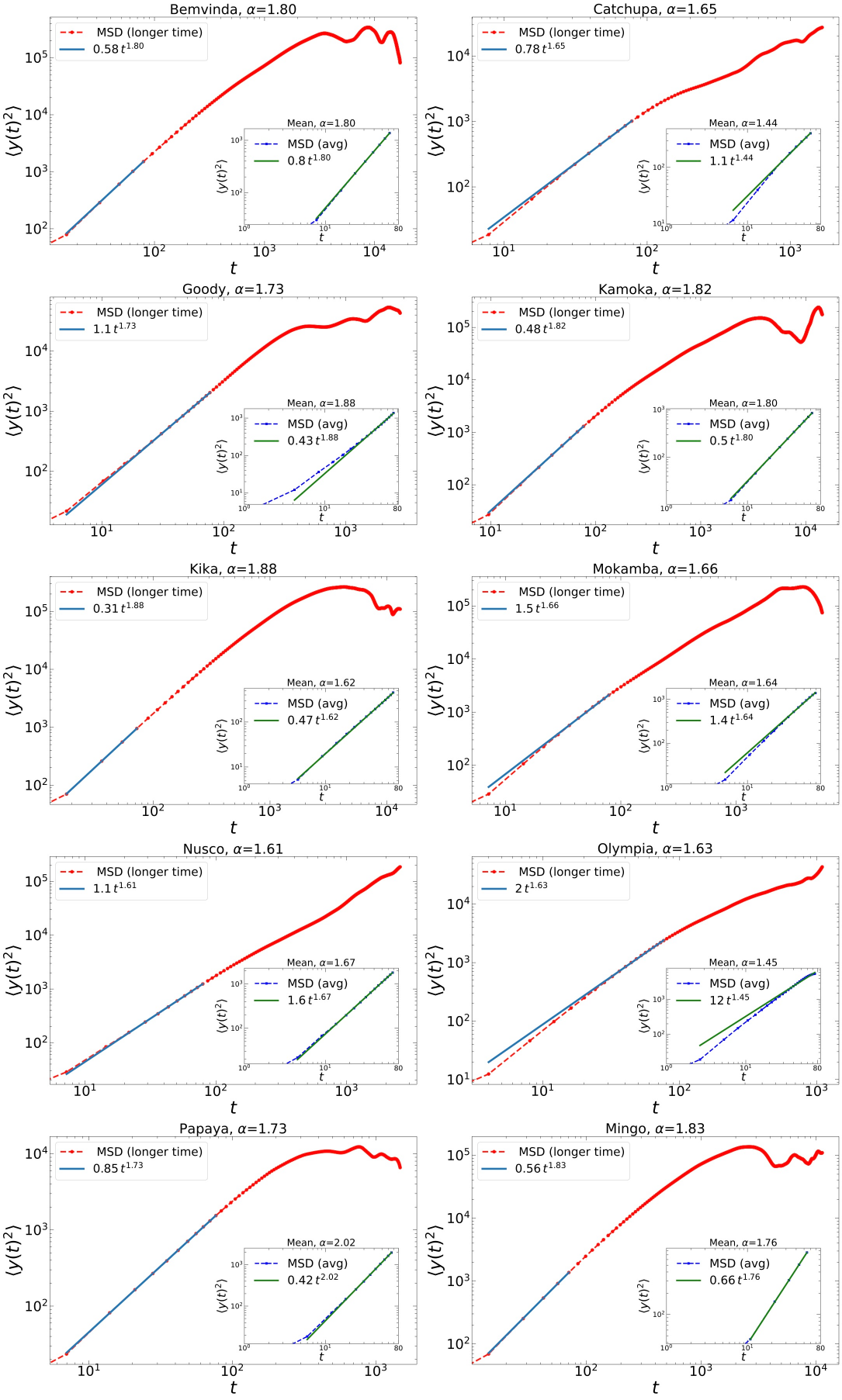} 
    \caption{TAMSD for $\langle y(t)^2 \rangle$. The motion all turtles is superdiffusive in the $y$ direction at times up to hundreds of hours. The displacement $y$ is measured in km.}
    \label{fig:FigS17}
\end{figure}

The time averaged mean-squared displacement (TAMSD) can be computed by
\begin{equation}
\langle M^2(\tau)\rangle=\frac{1}{N-\tau}\sum_{i=1}^{N-\tau}(X(i+\tau)-X(i))^2.
\end{equation}
We computed it for the following two cases of
(i) the full longest path and (ii) the ensemble of trajectory segments obtained after removing the largest gaps of time intervals, as explained above.
For the trajectory segments, we first considered trajectories with a given time duration by computing the TAMSD for each of these, and then ensemble averaged over these TAMSDs.
As one can see from Figs. \ref{fig:FigS16},\ref{fig:FigS17} the motion of all turtles is superdiffusive on a scale up to hundreds of hours.
This conclusion also holds when we analyse the MSD for the full trajectories (i.e. including the gaps). We remark that here we have truncated the noisy tails of the MSDs, which emerge from time averaging along the full trajectories by running out of data for larger time lags along the trajectory. Interestingly, the superdiffusive time scale corresponds to the oscillation period of the velocity autocorrelation function, as we show in detail below.
At these short times we fit the TAMSD with the power law function $at^{\alpha}$. For all the turtles, the exponent $\alpha$ lies in the interval
$1<\alpha<2$.

\subsection{Analysis of the turning angles}
\begin{figure}[H]
    \centering
     \includegraphics[width=\columnwidth]{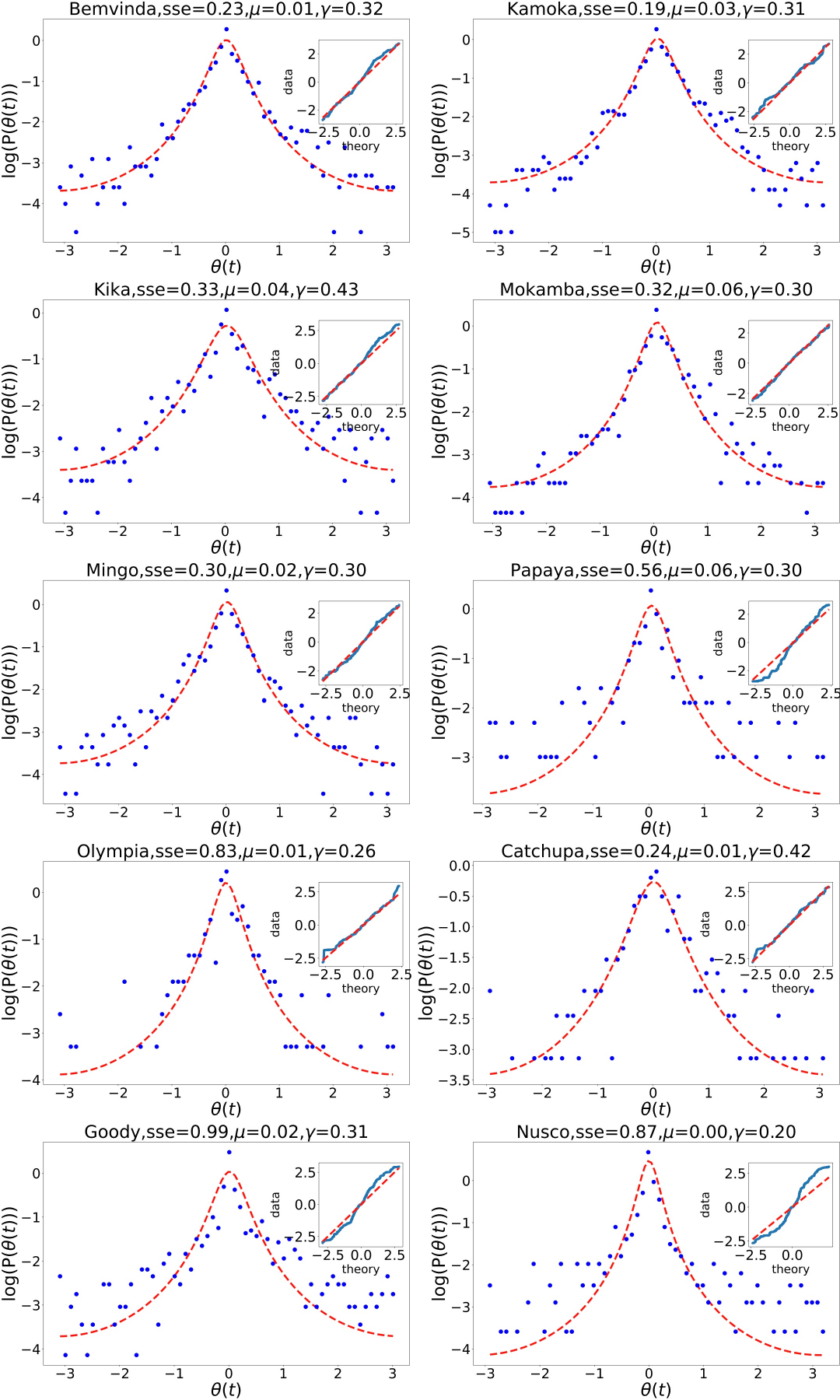} 
    \caption{Log-linear plot for the turning angles. Their distributions are well fitted by the Cauchy distribution. More importantly, we do not see any chirality, as the turning angle distribution is always centred around zero. This indicates that the turtle motion differs from the well studied chiral active motion, yet, it displays loops and oscillatory VACFs. The turning angles are measured in radian.}
    \label{fig:FigS18}
\end{figure}

The orientation angle $\phi$ can be computed with respect to the fixed Cartesian frame as follows:
$$\phi=\tan^{-1}\left(\frac{v_y}{v_x}\right).$$
The turning angle $\Delta \phi =\theta$ is then computed as the change in the orientational angle:
$$ \theta=\phi(t+\Delta t )-\phi(t).$$
We find that the turning angle distribution is well fitted by a wrapped Cauchy distribution which is given by following functional form:
$$f(\theta, \mu , \gamma)=\frac{1}{2\gamma} \frac{\sinh \gamma}{\cosh \gamma-\cos(\theta-\mu) }.$$
In the inset, the Q-Q plot shows strong agreement with this distribution.
As we can see in Fig. \ref{fig:FigS18}, there is no chirality since the mean $\mu \approx 0$.
We have used third stage cleaned data to perform the turning angle analysis. Due to the linear interpolation, this yielded a disproportionate amount of angles with zero values. Accordingly, we have appropriately truncated the resulting large peak at zero in the distributions. We remark that for time series with irregularly-spaced time intervals there is a problem to reliably obtain the turning angle from the data. Here our focus is not on the precise details of the turning angle distribution but rather on its symmetry, which is not affected by the above particular analysis.

\subsection{Velocity autocorrelation function}
The velocity autocorrelation function (VACF) provides important information about the non-Makovian nature of a stochastic process, as well as topological features of the corresponding trajectories, for instance, the presence of loops. It is a natural measure to test for memory in the movements of organisms.

\begin{figure}[H]
    \centering
     \includegraphics[width=\columnwidth]{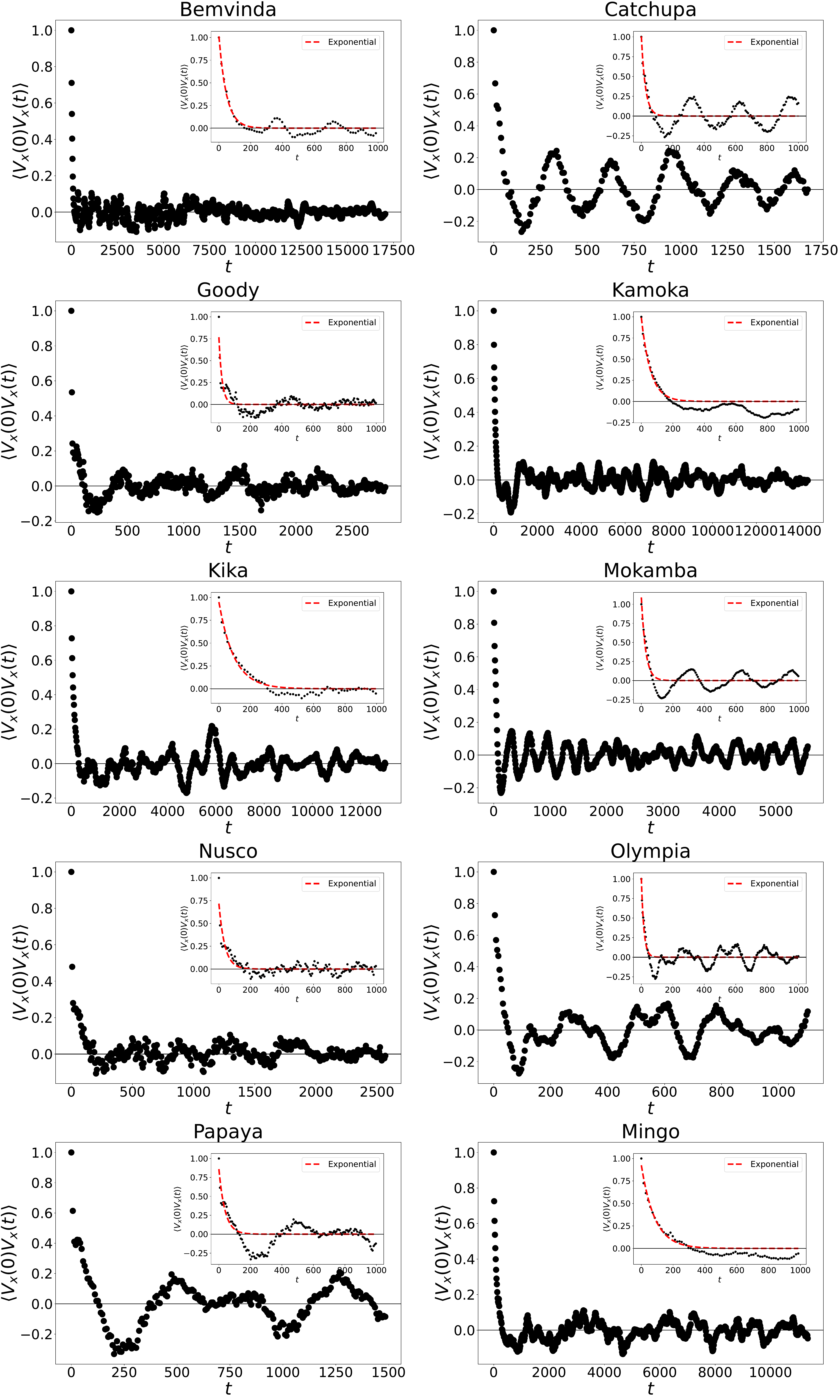} 
    \caption{Normalised velocity autocorrelation function (VACF) for the $x$ component  $v_{x}$ of the turtle data. The unit of the velocity is km/h. Times are shown in hours. In the inset, we fit the VACF at short times with an exponential function. As we see from the inset, it shows distinctive persistent oscillations for several turtles including Mokamba, Olympia, Catchupa, Papaya and Goody. }
    \label{fig:FigS19}
\end{figure}

\begin{figure}[H]
    \centering
     \includegraphics[width=\columnwidth]{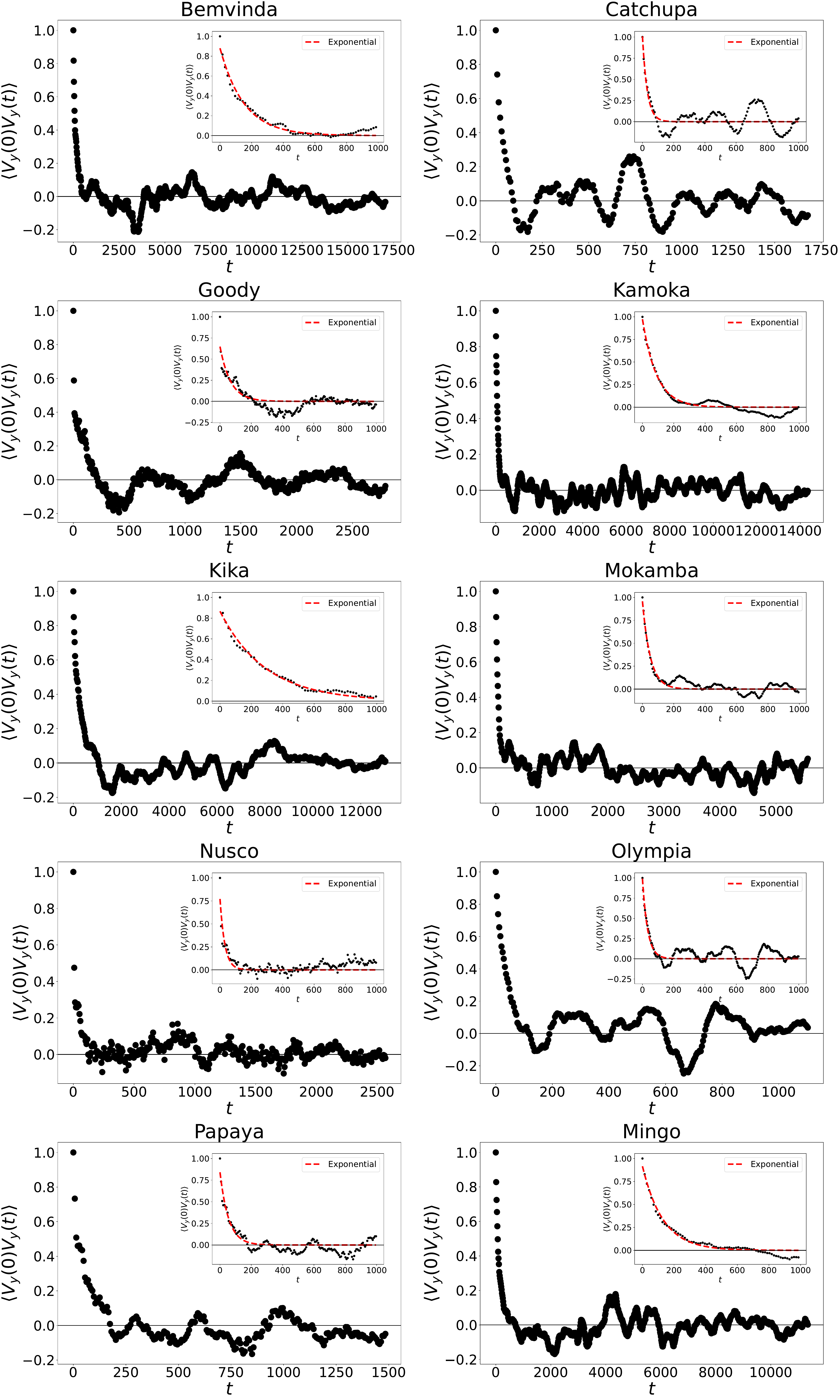} 
    \caption{Normalised velocity autocorrelation function (VACF) for the $y$ component  $v_{y}$ of the turtle data. The unit of the velocity is km/h. Times are shown in hours. }
    \label{fig:FigS20}
\end{figure}

In our case, we define the VACF as a time average along the path of a turtle:
\begin{equation}
    C(\tau)=\frac{1}{m} \sum_{i=1}^{m} v(t_i)v(t_i+\tau).
\end{equation}
For the experimental data we compute the normalised VACF. Fig. \ref{fig:FigS19} and Fig. \ref{fig:FigS20} show the normalised  VACFs $\langle V_{x} V_{y} \rangle$ and $\langle V_{x} V_{y} \rangle$ (we use capital letters for the normalised VACF) for all turtles.  We have truncated the noisy tails for the VACFs.
We fit these VACFs with an exponential function at short times as shown in the insets of Fig. \ref{fig:FigS19} and Fig. \ref{fig:FigS20}.
At longer times, we observe persistent oscillations. The periodicity of the oscillations is particularly pronounced for Mokamba, Catchupa, Papaya and Olympia and Goody. For Kamoka and Mingo, we do not observe periodic oscillations because of the spatio-temporal mixing of several loops, whereas for the turtles Bemvinda and Kika, there are long gaps. Therefore, we exclude the data-driven stochastic analysis for the rest of the turtles.

\subsection{Velocity cross-correlation}

The empirical velocity cross-correlation is evaluated using the mean-subtracted velocity fluctuations, defined as $\delta v_{x}(t) = v_{x}(t) - \langle v_{x} \rangle$ and $\delta v_{y}(t) = v_{y}(t) - \langle v_{y} \rangle$, where $\langle v_{x} \rangle$ and $\langle v_{y} \rangle$ denote the mean velocities over the trajectory.
\begin{equation}
    C_{V_x V_y}(\tau) = \frac{ \sum_{t} \delta v_x(t+\tau) \cdot \delta v_y(t) }{ \sqrt{ \sum_{t} \left(\delta v_x(t)\right)^2 \cdot \sum_{t} \left(\delta v_y(t)\right)^2 } }
\end{equation}
and
\begin{equation}
    C_{V_y V_x}(\tau) = \frac{ \sum_{t} \delta v_y(t+\tau) \cdot \delta v_x(t) }{ \sqrt{ \sum_{t} \left(\delta v_y(t)\right)^2 \cdot \sum_{t} \left(\delta v_x(t)\right)^2 } }.
\end{equation}
The normalized cross-correlation functions at a given time lag $\tau$ are computed using the \texttt{matplotlib.pyplot.xcorr} function~\cite{Hunter07} (with \texttt{normed=True}) which internally uses \texttt{numpy.correlate}~\cite{Harris20}.

\begin{figure}[H]
    \centering
      \includegraphics[width=\columnwidth]{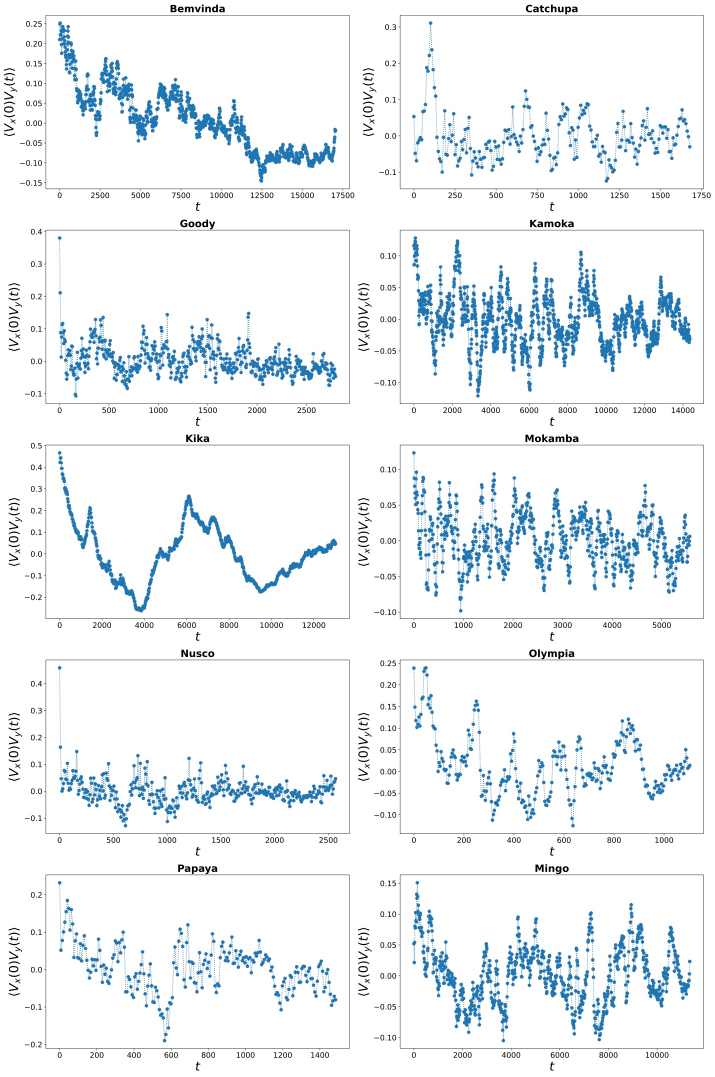} 
    \caption{Cross-correlation analysis of the velocity components $v_{x}$ and $v_{y}$ in terms of the normalized cross-correlation function $C_{V_x V_y}(\tau)$.}
    \label{fig:FigS21}
\end{figure}

Figure \ref{fig:FigS21} shows the normalized cross-correlations $C_{V_x V_y}(\tau)$ for all ten turtles. Results for $C_{V_y V_x}(\tau)$ look qualitatively similar, hence are not included here. In contrast to the regularly oscillating autocorrelations of $v_{x}$ shown before, for seven of the turtles the cross-correlations look somewhat randomly fluctuating. For three out of the ten turtles, however, they look a bit different: In the case of Kika and Bemvinda, there are very large gaps in the trajectories. As we linearly interpolate over these gaps for computing correlation functions, it is not too surprising that these gaps lead to large-scale trends in the data, as visible in the results for both turtles. Goody also seems to be a slightly special case, as here one can see somewhat regular, though noisy oscillations with a period that even roughly matches the one in the $v_x$ autocorrelations. Note, however, that Goody performs loops over very large spatial scales, in contrast to the other turtles, which might explain these slightly deviating results. This demonstrates again a certain individuality of the turtles, plus the limitations of our analysis due to the available data sets. In general, we conclude the absence of correlations between $v_{x}$ and $v_{y}$, which in turn rules out odd diffusion.

\subsection{Impact of the day night cycle}
\begin{figure}[H]
    \centering
     \includegraphics[width=\linewidth]{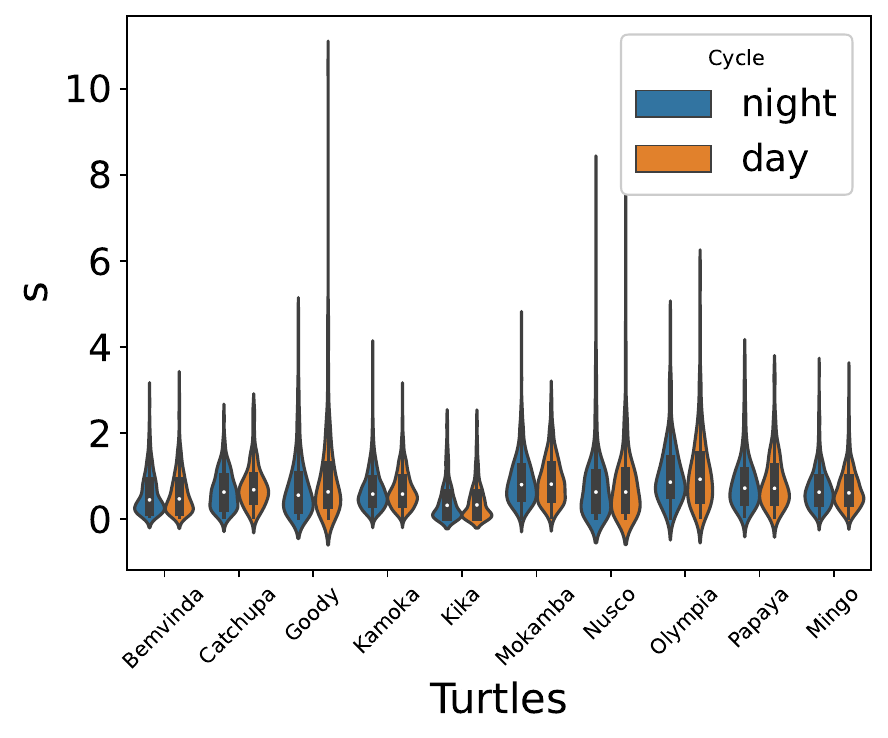} 
    \caption{Violin plot for the speed data of all turtles ($s$ is in km/h). We do not observe any impact of day night cycle on the speed distribution.}
    \label{fig:FigS22}
\end{figure}

\begin{figure}[H]
    \centering
     \includegraphics[width=\linewidth]{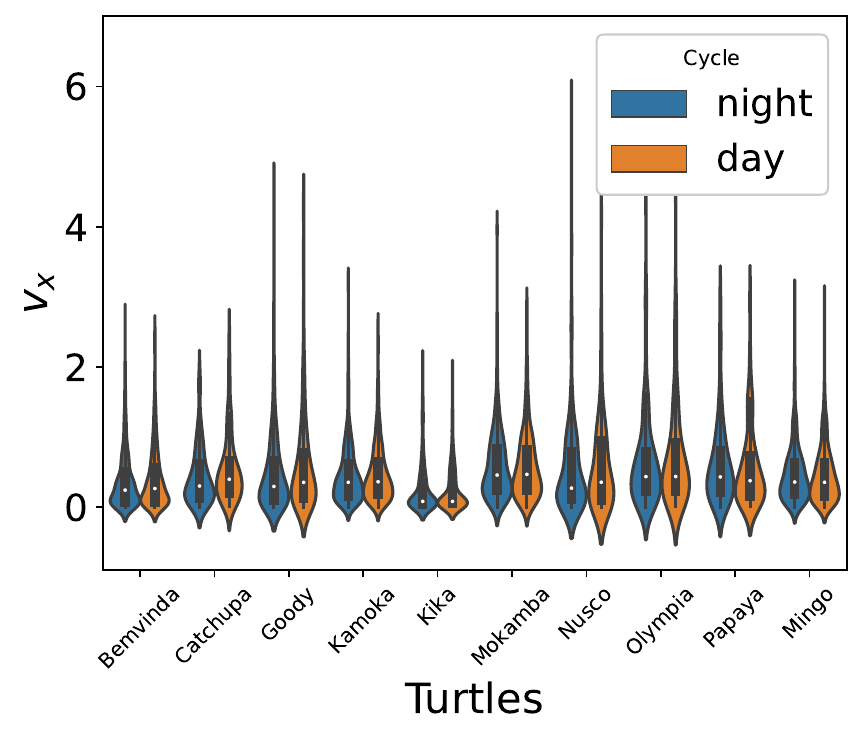} 
    \caption{Violin plot for $v_{x}$(measured in km/h).}.
    \label{fig:FigS23}
\end{figure}

\begin{figure}[H]
    \centering
     \includegraphics[width=\linewidth]{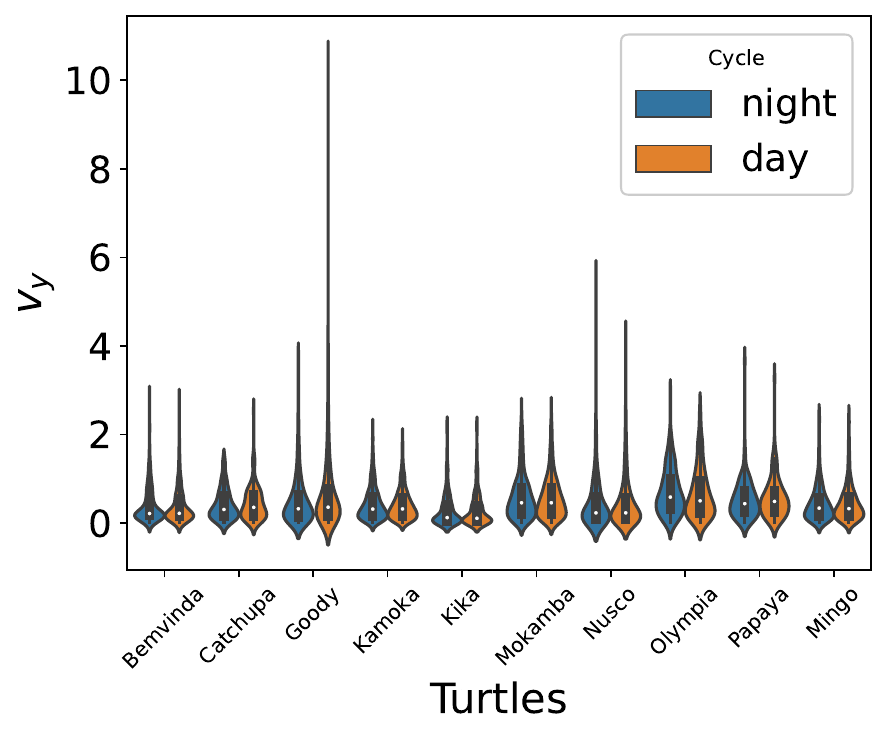} 
    \caption{Violin plot for $v_{y}$(km/h).}.
    \label{fig:FigS24}
\end{figure}

We also analysed the impact of the day-night cycle on the speed $s$ as well as the velocities $v_{x}$, $v_{y}$. As we can observe from the violin plots (Figs. \ref{fig:FigS22},\ref{fig:FigS23},\ref{fig:FigS24}), we do not see any impact on speed or velocities. This is consistent with the fact that the sleep duration of the turtle is less than the "average" interval over which we measure the speed. We have also plotted the normalised histogram and, as it turns out, both day as well as night speed distributions align well and we do not see any qualitative difference. This shows that the deviation of the tail part of speed distributions from a Rayleigh distribution is not caused by the day-night cycle.

\section{Data driven stochastic modelling}  
\subsection{Introduction: the overdamped stochastic model}
Our statistical data analysis presented above shows that turtles typically move with velocities that are Gaussian in the centre, with exponential tails, corresponding to speed distributions that are Rayleigh in the centre, with exponential tails. The associated VACFs are typically exponentially decaying for short times with, interestingly and importantly, non-trivial, often persistent oscillations on long time scales. We have shown above that these persistent oscillations are not due to chirality, nor do they reflect odd diffusion. Hence, in the following we construct a novel stochastic movement
from this data, for which we directly feed in these central results. This model is motivated by respective previous works on stochastically modelling bumblebee flights \cite{LICCK12,LCK13} and cell migration \cite{Diet22}. It is also motivated within the more general context of active Brownian motion in view of one known active particle model known as the active Ornstein-Uhlenbeck process, see \cite{KLK26} for a review.

Accordingly, we use as a starting point the simple overdamped Langevin equation
\begin{gather*}
    \dot{x}=\eta(t),\\ 
     \dot{y}=\eta(t)
\end{gather*} 
by taking our experimental observations referred to above as an input. Using the fitted functional forms of the VACFs with persistent oscillations and approximations of the obtained velocity distributions, we will show that the model generates superdiffusion on an intermediate time scale as well as looping. Since we have verified that cross-correlation between the $x$ and $y$ directions is negligible, we formulate our model without coupling between $x$ and $y$ positions. 

\subsection{Procedure for the data driven modelling}

In the following we present the procedure that we implemented for stochastically modelling the data of the turtle Mokamba:

\begin{itemize} 
    \item Generate uncorrelated samples from a desired probability distribution: In our case, we chose to ignore the deviations in the tail and generated Gaussian samples of the velocities.
    \item Use the fitted VACF as an input for the covariance matrix of the noise. 
    \item Then use Cholesky decomposition for the covariance matrix to generate the desired velocities, as follows:
    
The Cholesky decomposition for a positive definite  square matrix A is defined as 
\[
A = LL^T = 
\begin{bmatrix}
L_{11} & 0 & 0 \\
L_{21} & L_{22} & 0 \\
L_{31} & L_{32} & L_{33}
\end{bmatrix}
\begin{bmatrix}
L_{11} & L_{21} & L_{31} \\
0 & L_{22} & L_{32} \\
0 & 0 & L_{33}
\end{bmatrix}.
\]       
Subsequently, the individual components of the lower triangular matrix L can be obtained through the following equations:  
\begin{equation}
L_{jj} = \sqrt{A_{jj} - \sum_{k=1}^{j-1} L_{jk}^2},
\end{equation}
\begin{equation}
L_{ij} = \frac{1}{L_{jj}} \left( A_{ij} - \sum_{k=1}^{j-1} L_{ik}L_{jk} \right) \quad \text{for } i > j.
\end{equation}
To generate a random vector $\mathbf{X}$ with a specific covariance matrix $\Sigma$, we can then use Cholesky decomposition $\Sigma = LL^T$. 
Given a vector of independent samples $\mathbf{Z} \sim \mathcal{N}(0, I)$, the transformed vector is:
\[
\mathbf{X} = \mu + L\mathbf{Z}.
\] 
The covariance of $\mathbf{X}$ is then verified as:
\begin{align*}
\text{Var}(\mathbf{X}) &= \text{Var}(L\mathbf{Z}) = L \text{Var}(\mathbf{Z}) L^T = L I L^T = LL^T = \Sigma.
\end{align*} 
In our case, the random vector  $\mathbf{X}$  is the velocity  and  the covariance matrix $\Sigma $  is defined through the VACF.    
    \item Implement boundary conditions as forces to incorporate the effect of the environmental constraints: The coastline can be treated as a steep potential while the deep sea boundary is softer, hence we used an asymmetric harmonic potential. 
    \item Parameter selection: we chose suitable parameter values to match the experimental results.    
\end{itemize} 

\subsection{Implementation of boundary conditions}
\begin{figure}[H]
     \includegraphics[width=\linewidth, height=0.70\textheight, keepaspectratio]{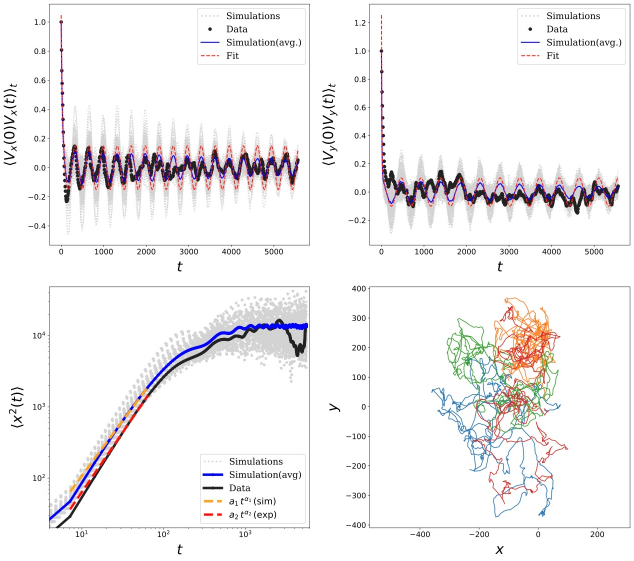} 
    \caption{Stochastic data-driven modelling for Mokamba. We used Cholesky decomposition to construct colored noise consistent with non-Markovian memory effects (i.e., an exponentially decaying VACF at short times with persistent  oscillations for longer times). The boundaries along $x$ are included as an asymmetric harmonic potential. The model reproduces superdiffusive motion as well as the spatial asymmetry of the Mokamba trajectory. The time $t$ is measured in hours, the displacement $x$ is measured in km, the velocities are measured in km/h.}
    \label{fig:FigS25} 
\end{figure}

\begin{figure}[H]
    \centering
     \includegraphics[width=\linewidth, height=0.70\textheight, keepaspectratio]{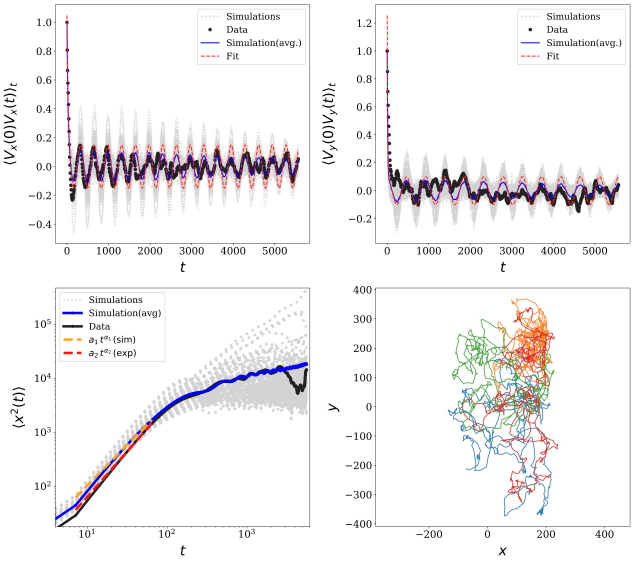} 
    \caption{Corresponding results for Mokamba for LJ potential boundaries along $x$. As we can see, here the MSD does not attain a plateau (bottom left), even though this is difficult to see in the trajectories (bottom right).}
    \label{fig:FigS26}
\end{figure}

The unbounded motion replicated loops and the superdiffusive MSD at shorter times. However, it could not reproduce the asymmetry of the Mokamba trajectory as well as the saturation (plateau) of the MSD reached for the $x$-direction. Accordingly, we tried several boundary conditions for the $x$-direction including hard reflective boundaries, symmetric and asymmetric harmonic potentials, as well as the Lennard-Jones (LJ) potential. Based on this analysis, we found that an asymmetric harmonic potential is better suited to reproduce the movements of Mokamba than the LJ potential.   
We used an asymmetric harmonic potential with different stiffness values, loosely capturing the effect of the coastline on the right-hand side ($x>0$, measured in km) with a steeper potential ($k_2 = 0.18\,\mathrm{h}^{-1}$) and a smoother potential (with stiffness parameter $k_1 = 0.01\,\mathrm{h}^{-1}$) on the left-hand side ($x\leq 0$). The potential is defined as

\begin{equation}
V(x) =
\begin{cases}
\frac{1}{2}k_{1}\,x^{2}, & x \le 0,\\[4pt]
\frac{1}{2}k_{2}\,x^{2}, & x > 0.
\end{cases}
\end{equation}
We include these boundary conditions in the overdamped Langevin equation given above as force terms, leading together with the noise term to the two-dimensional generalised Langevin equation
 \begin{equation}
d{\bf x}/dt={\bf F}({\bf x})+\boldsymbol{\zeta}(t).      
 \end{equation}
The following two figures show the results from solving this stochastic model numerically for Mokamba compared to the experimental data, see Fig. \ref{fig:FigS25} for the asymmetric harmonic and Fig. \ref{fig:FigS26} for the LJ potential.
\subsection{Modelling other turtles} 
We implemented this data-driven stochastic modelling framework as much as possible for the other turtles, i.e., as far as VACFs could be extracted reliably from the experimental data, namely for Catchupa, Goody, Olympia, Papaya and Goody. 
The details of the computed exponents of the MSD are tabulated further below in Table \ref{tab:scale_matching}. 

\begin{figure}[H]
    \centering 
     \includegraphics[width=\linewidth, height=0.70\textheight, keepaspectratio]{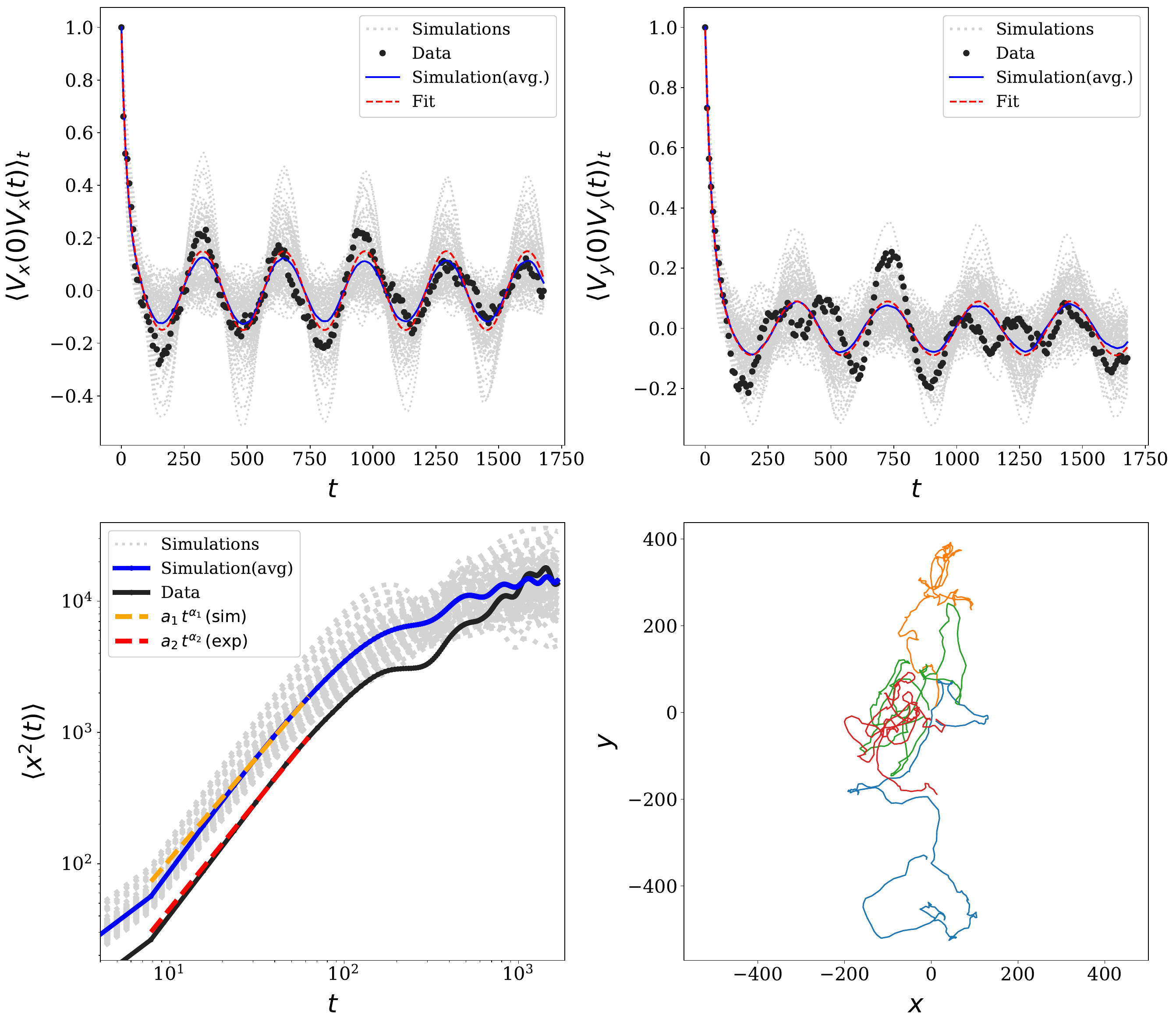} 
    \caption{Ensemble average results for Catchupa with harmonic potential. The stiffness for  $x \leq 0 $  is $ k_1 = 0.01\,\mathrm{h}^{-1}$ and for $x>0$, $k_2=0.06\, \mathrm{h}^{-1} $}.
    \label{fig:FigS27}
\end{figure} 

\begin{figure}[H]
    \centering
     \includegraphics[width=\linewidth, height=0.70\textheight, keepaspectratio]{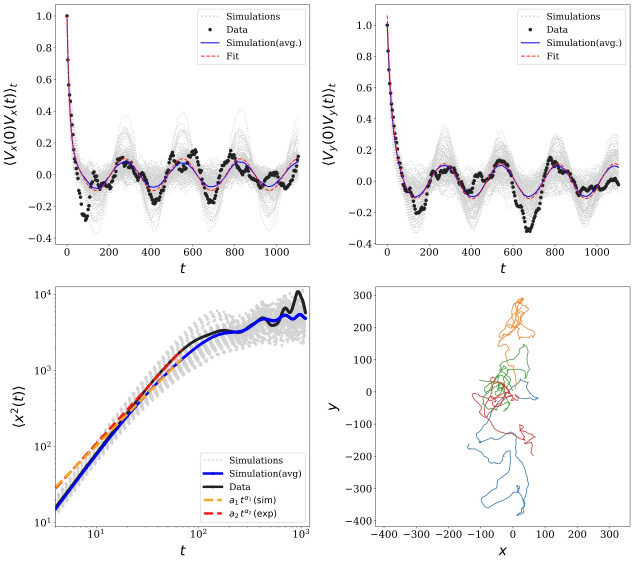} 
    \caption{Ensemble average results for Olympia with harmonic potential. The stiffness for  $x \leq 0 $  is $ k_1 = 0.01\,\mathrm{h}^{-1}$ and for $x>0$, $k_2=0.06\, \mathrm{h}^{-1} $}.
    \label{fig:FigS28}    
\end{figure}

\begin{figure}[H]
    \centering
     \includegraphics[width=\linewidth, height=0.70\textheight, keepaspectratio]{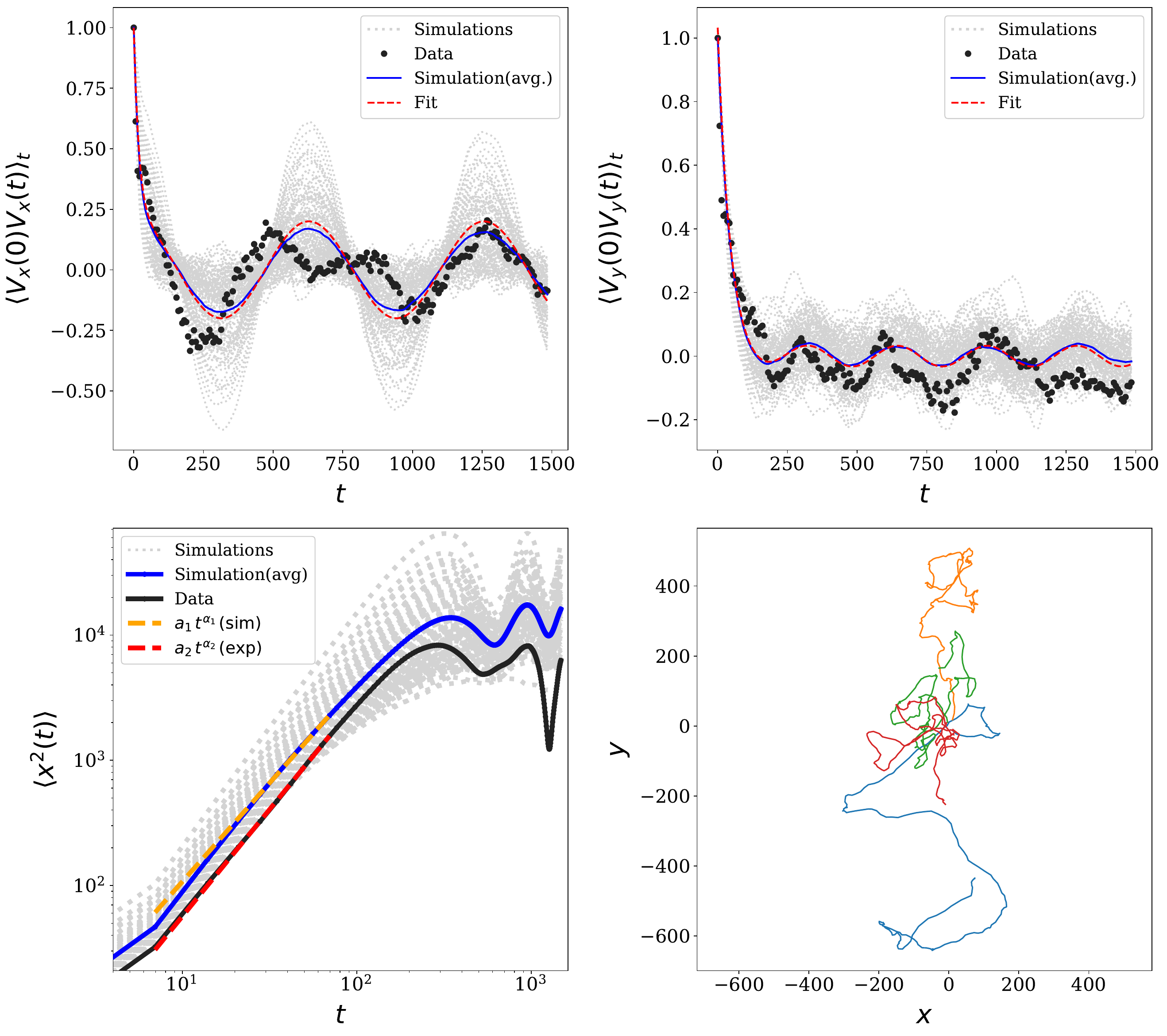} 
    \caption{Ensemble average results for Papaya with harmonic potential. The stiffness for  $x \leq 0 $  is $ k_1 = 0.01\,\mathrm{h}^{-1}$ and for $x>0$, $k_2=0.06\, \mathrm{h}^{-1} $}.
    \label{fig:FigS29}
\end{figure}

\begin{figure}[H]
    \centering
     \includegraphics[width=\linewidth, height=0.70\textheight, keepaspectratio]{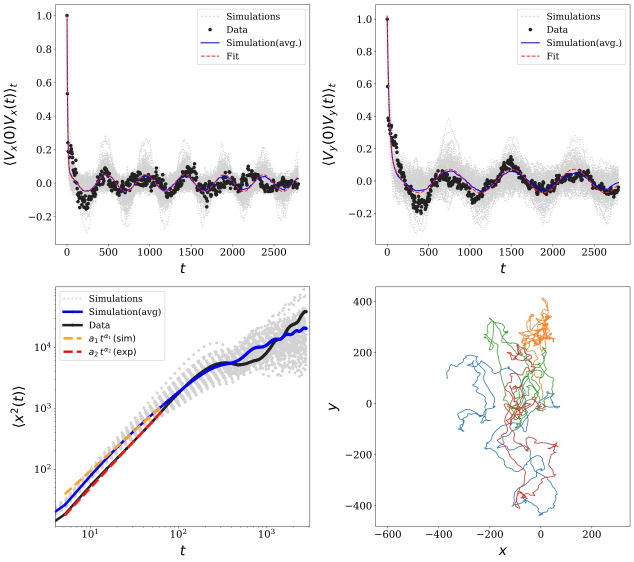} 
    \caption{Ensemble average results for Goody with harmonic potential. The stiffness for  $x \leq 0 $  is $ k_1 = 0.001\,\mathrm{h}^{-1}$ and for $x>0$, $k_2=0.06\, \mathrm{h}^{-1} $}.
    \label{fig:FigS30}
\end{figure}

\subsection{Simulations of a tracer particle moved by the oceanic currents}

An obvious question is whether the loops in the turtle movements that we observed above are generated by ocean currents, as observed in Ref. \cite{LLL06} for juvenile leatherback sea turtles swimming in front of South Africa.
In order to check for the impact of such currents we performed Lagrangian tracer simulations, using
the OceanParcels v2.2.2 library in Python 3.0 (virtual hatchling). More details can be found in Ref. \cite{ThesisRT2022}.
The  gray lines in Fig.1 of the main text display results for the motion of 50 tracer particles. 
As we can see, they follow complete different trajectories compared to the turtle Mokamba. This rules out that the loops displayed by Mokamba are caused by ocean currents.   

\section{Spatio-temporal scale matching}

We now clarify the connection between the large-scale loops performed by the turtles, characterised by the oscillatory VACFs, and the superdiffusion on intermediate time scales observed in the MSDs. The basic idea is to calculate the time scale of the persistent oscillations from a given VACF, and to relate this time scale to the functional form of the MSD. We can then even obtain a mean distance travelled up to this time scale from the MSD and check for an environmental coupling of the turtle motion related to these spatio-temporal scales inherent in the turtle motion. We call this procedure spatio-temporal scale matching.

In the following we demonstrate this approach for the motion of the turtle in the $x$ direction. First, the characteristic temporal duration of a half-cycle in the VACF for $v_x$, $T_{x,1/2}$, was obtained by halving the mean time between local minima in $\langle v_x(0)v_x(\tau)\rangle$. This half-period $T_{x,1/2}$ was estimated by two different methods:

\paragraph{Method 1: Fitted VACF.} 
The VACF of $v_x$ was first fitted with a function representing the sum of an exponential decay and a cosine wave
$C(t) =A \exp(-Bt) + C \cos(Dt)$ using \texttt{lmfit} where A, B, C and D are real constants.  Local minima were computed from the \emph{fitted} VACF. Then $T_{x,1/2}^{\mathrm{fit}}$ was taken as half of the mean spacing between consecutive minima.
\paragraph{Method 2: Raw VACF.} 
Local minima were detected directly on the \emph{empirical} VACF computed from the data via \texttt{numpy.acorr}. $T_{x,1/2}^{\mathrm{raw}}$ is again half the mean inter-minima spacing, but now based on the raw signal. This is a more model-independent estimate. 
In both cases, $\sqrt{\langle x^2(T_{x,1/2})\rangle_T}$ is then evaluated from a power-law fit to the experimental MSD  computed at $T_{x,1/2}$. In the following we first show results by applying method one, then for method two.
\subsection{Scale matching using the fitted velocity autocorrelations}
\label{sec:sm_fitted} 
We apply the scale-matching procedure to the same turtles for which we constructed stochastic models. The following figures show the turtle path obtained from the experimental data (left), including an approximate scale for the core foraging region of the respective turtle. In the middle the VACF along $x$ extracted from the experimental data plus the fit are shown. The figure to the right displays the corresponding MSD, including the spatio-temporal scales identified by the scale matching approach. 

\begin{figure}[H]
    \centering
    \makebox[\textwidth][c]{ \includegraphics[width=\linewidth, height=0.70\textheight, keepaspectratio]{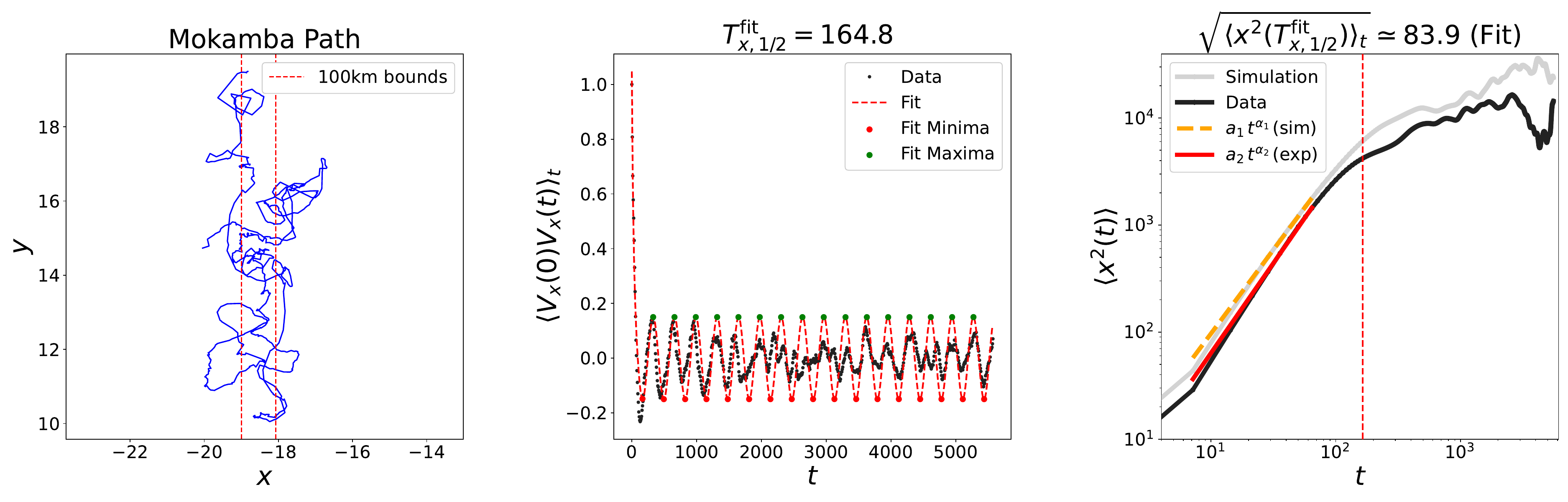} }    
    \caption{Single-trajectory scale matching for \textbf{Mokamba} using $T_{x,1/2}^{\mathrm{fit}}$ from the fitted VACF.
    \emph{Left}: Turtle path with longitudinal bounds displaced by $100\,$km.
    \emph{Middle}: Empirical $v_x$ VACF; the title shows $T_{x,1/2}^{\mathrm{fit}}$. 
   \emph{Right}: MSD from a single Cholesky simulation run (grey) vs. experimental data (black) with power-law fits; the red vertical marker indicates $T_{x,1/2}^{\mathrm{fit}}$.} 
    \label{fig:FigS31}
\end{figure}  

\begin{figure}[H]
    \centering
    \makebox[\textwidth][c]{ \includegraphics[width=\linewidth, height=0.70\textheight, keepaspectratio]{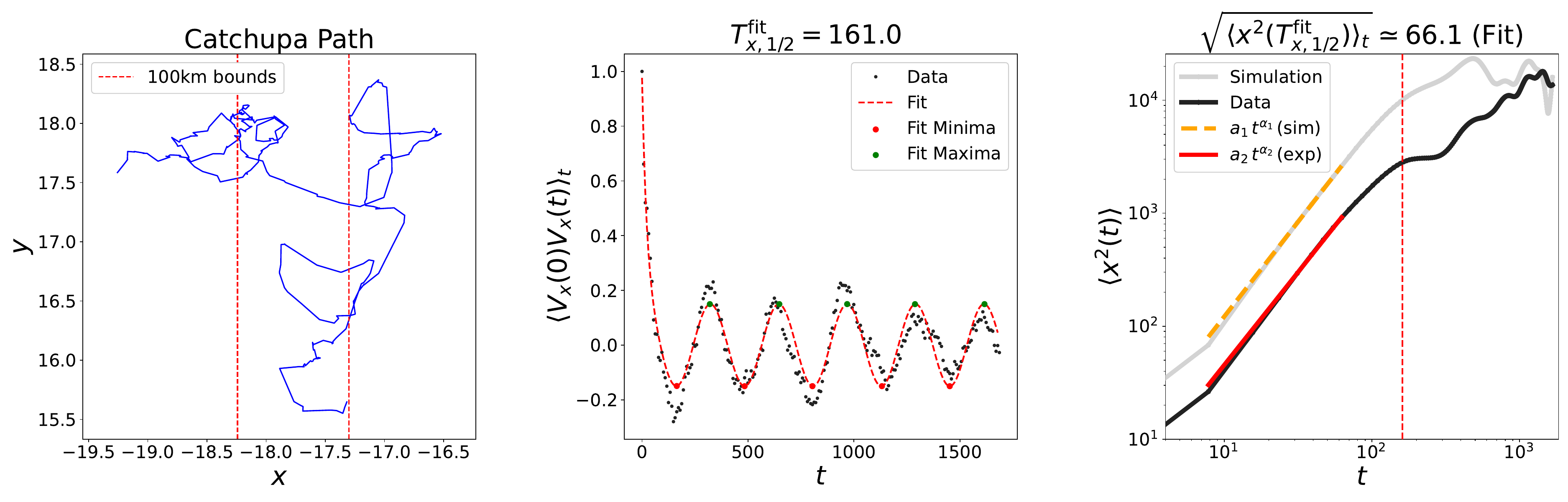} }
    \caption{Single-trajectory scale matching for \textbf{Catchupa} using $T_{x,1/2}^{\mathrm{fit}}$ from the fitted VACF. }
    \label{fig:FigS32}
\end{figure}

\begin{figure}[H]
    \centering
    \makebox[\textwidth][c]{ \includegraphics[width=\linewidth, height=0.70\textheight, keepaspectratio]{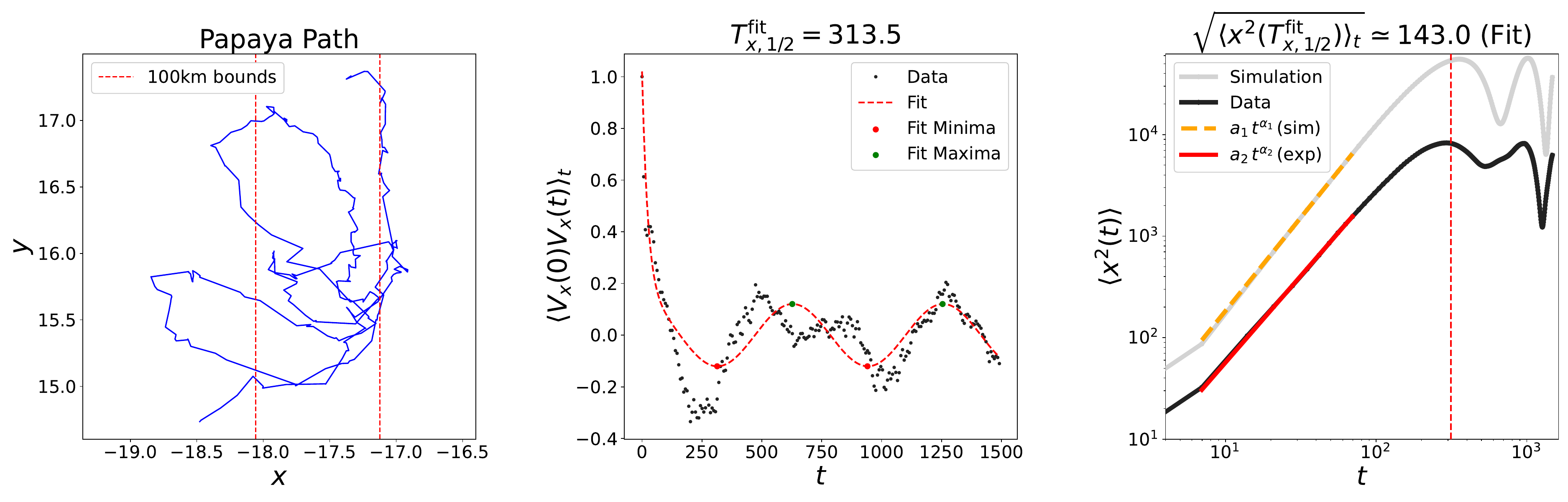} }
    \caption{Single-trajectory scale matching for \textbf{Papaya} using $T_{x,1/2}^{\mathrm{fit}}$ from the fitted VACF.}
    \label{fig:FigS33}
\end{figure}  

\begin{figure}[H]
    \centering
    \makebox[\textwidth][c]{ \includegraphics[width=\linewidth, height=0.70\textheight, keepaspectratio]{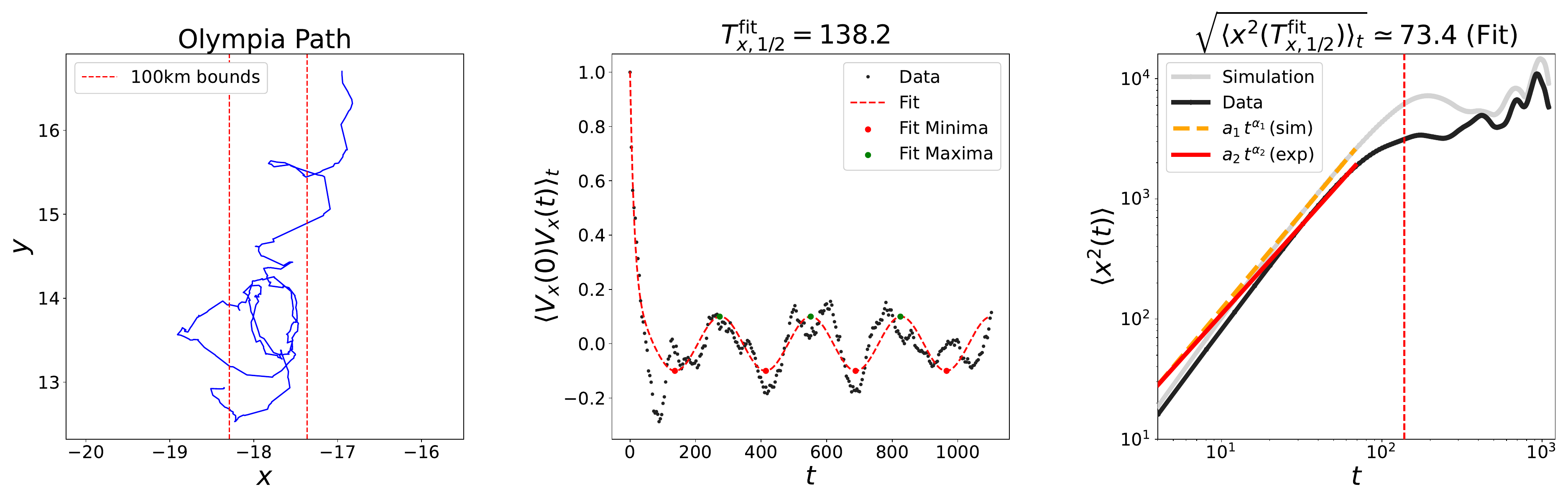} }
    \caption{Single-trajectory scale matching for \textbf{Olympia} using $T_{x,1/2}^{\mathrm{fit}}$ from the fitted VACF.}
    \label{fig:FigS34}
\end{figure} 

\begin{figure}[H]
    \centering
    \makebox[\textwidth][c]{ \includegraphics[width=\linewidth, height=0.70\textheight, keepaspectratio]{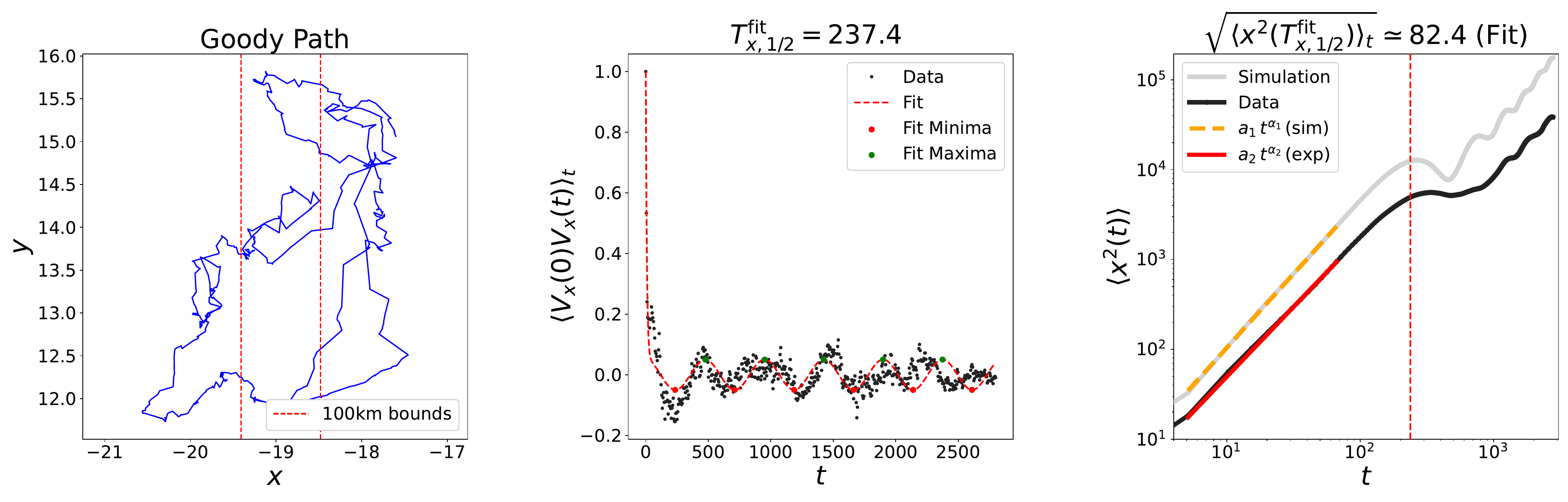} }
    \caption{Single-trajectory scale matching for \textbf{Goody} using $T_{x,1/2}^{\mathrm{fit}}$ from the fitted VACF.}
    \label{fig:FigS35}
\end{figure} 
\subsection{Scale matching using the raw velocity autocorrelations}
\label{sec:sm_raw}
The same way as in the previous section, here we show corresponding results by applying method 2, i.e., extracting the half period of the oscillations in the VACF along $x$ from the raw data.

\begin{figure}[H]
    \centering
    \makebox[\textwidth][c]{ \includegraphics[width=\linewidth, height=0.70\textheight, keepaspectratio]{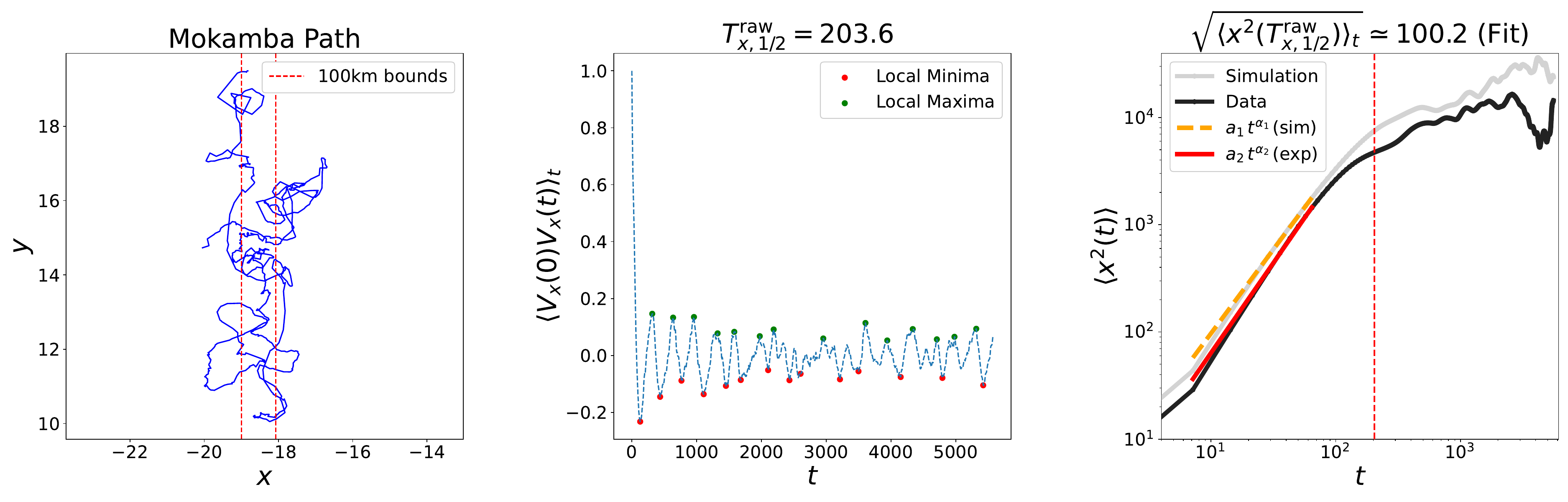} }
    \caption{Single-trajectory scale matching for \textbf{Mokamba} using $T_{x,1/2}^{\mathrm{raw}}$ from the raw VACF minima.
    \emph{Left}: Turtle path.
    \emph{Middle}: VACF with detected minima (red) and maxima (green); the title shows $T_{x,1/2}^{\mathrm{raw}}$. 
    \emph{Right}: MSD with scale-matching annotation derived from $T_{x,1/2}^{\mathrm{raw}}$. }
    \label{fig:FigS36}
\end{figure} 

\begin{figure}[H]
    \centering
    \makebox[\textwidth][c]{ \includegraphics[width=\linewidth, height=0.70\textheight, keepaspectratio]{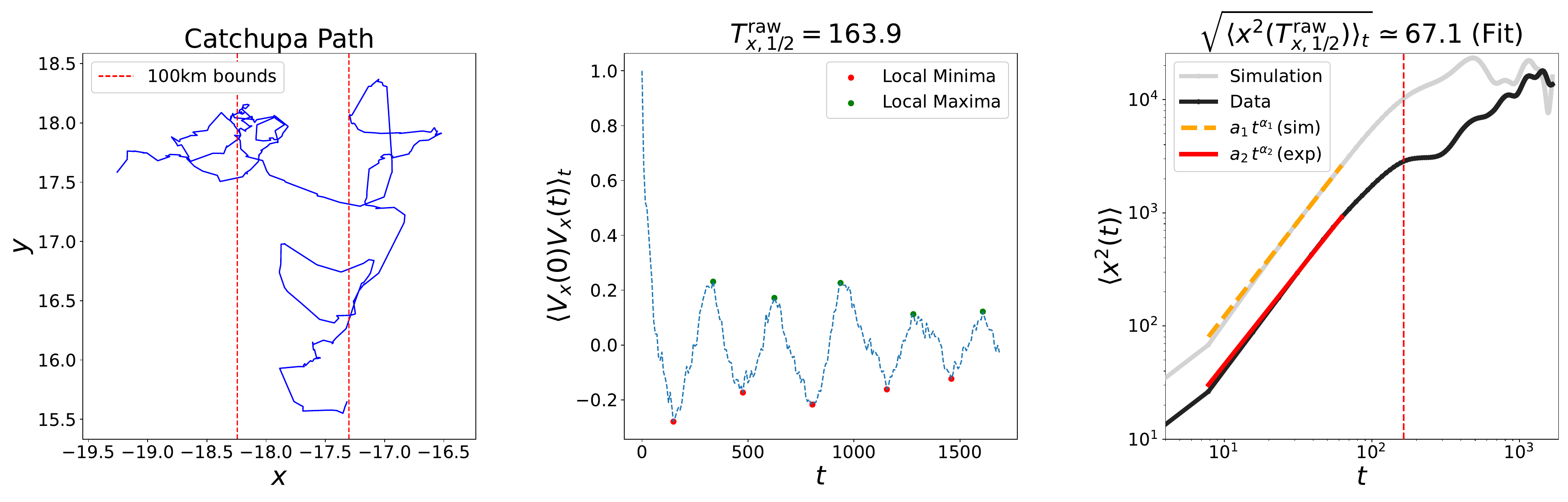} }
    \caption{Single-trajectory scale matching for \textbf{Catchupa} using $T_{x,1/2}^{\mathrm{raw}}$ from the raw VACF.}
    \label{fig:FigS37}
\end{figure}

\begin{figure}[H]
    \centering
    \makebox[\textwidth][c]{ \includegraphics[width=\linewidth, height=0.70\textheight, keepaspectratio]{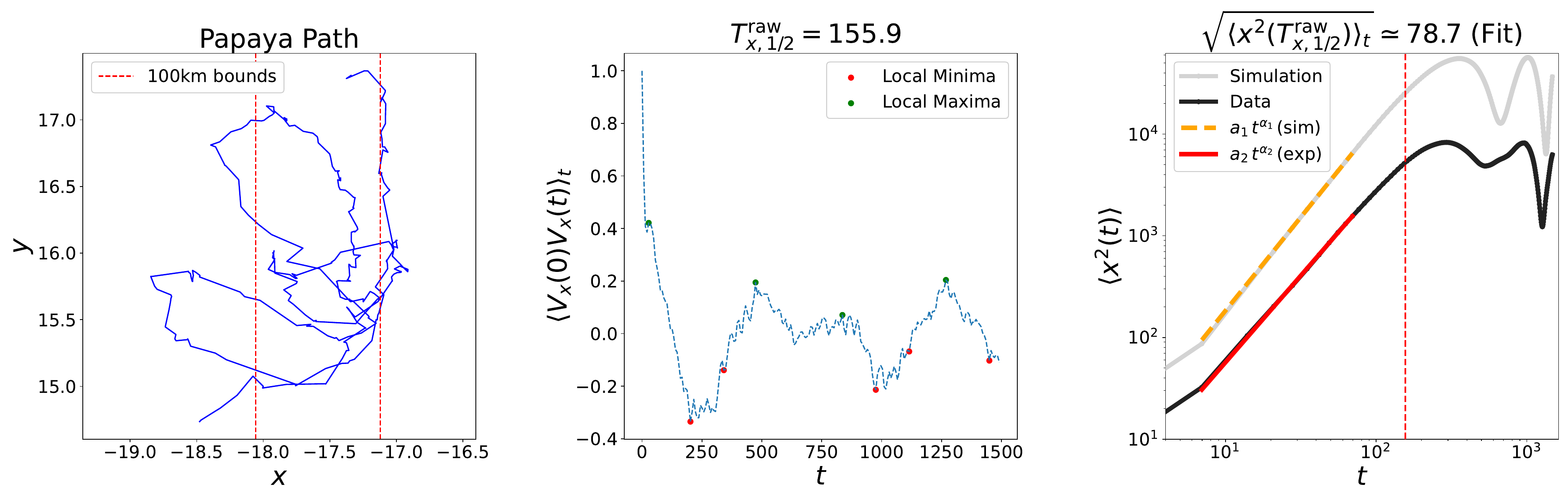} }
    \caption{Single-trajectory scale matching for \textbf{Papaya} using $T_{x,1/2}^{\mathrm{raw}}$ from the raw VACF.}
    \label{fig:FigS38}
\end{figure}   

\begin{figure}[H]
    \centering
    \makebox[\textwidth][c]{ \includegraphics[width=\linewidth, height=0.70\textheight, keepaspectratio]{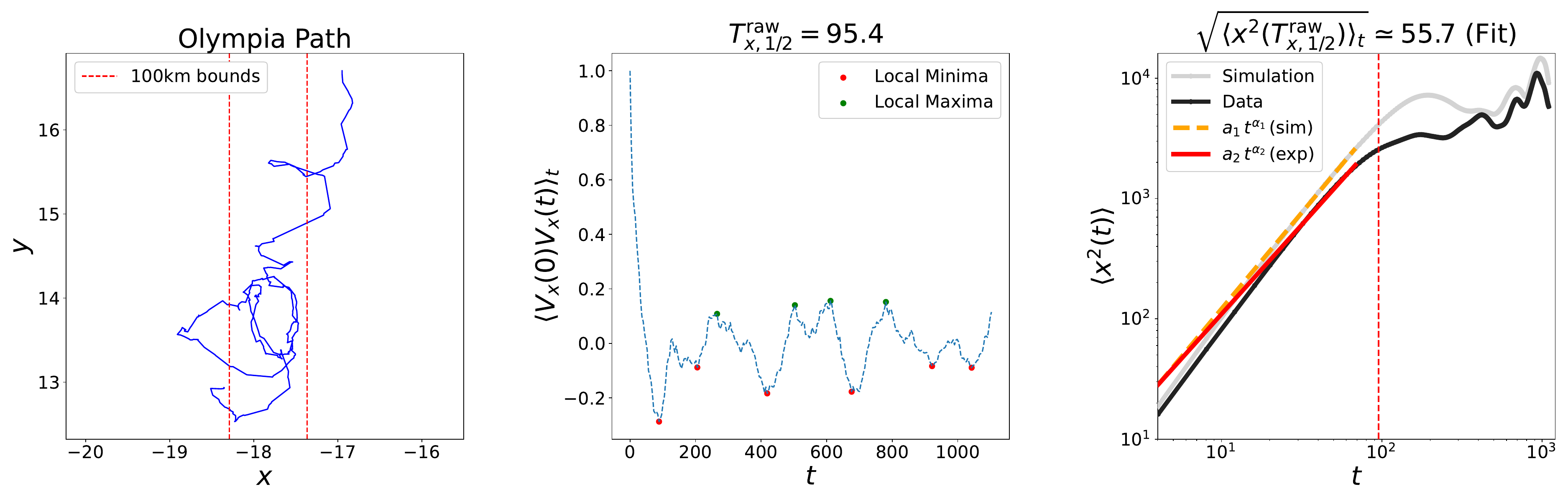} }
    \caption{Single-trajectory scale matching for \textbf{Olympia} using $T_{x,1/2}^{\mathrm{raw}}$ from the raw VACF.}
    \label{fig:FigS39}  
\end{figure}

\begin{figure}[H]
    \centering
    \makebox[\textwidth][c]{ \includegraphics[width=\linewidth, height=0.70\textheight, keepaspectratio]{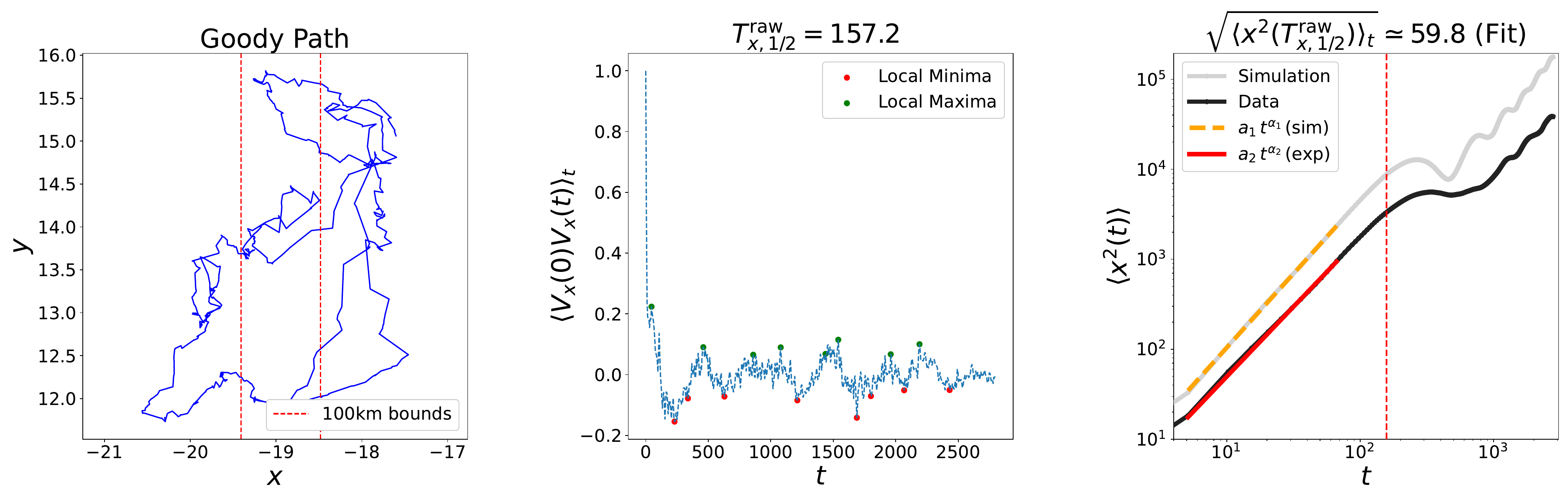} }
    \caption{Single-trajectory scale matching for \textbf{Goody} using $T_{x,1/2}^{\mathrm{raw}}$ from the raw VACF.}
    \label{fig:FigS40}  
\end{figure}       
\subsection{Summary}
\label{sec:sm_table}
The following Tab. \ref{tab:scale_matching} compares the two half-period $T_{x,1/2}$ estimates and the resulting
characteristic length scale $\sqrt{\langle x^2(T_{x,1/2})\rangle}$ for each turtle.
The MSD exponents $\alpha_{\mathrm{sim}}$ and $\alpha_{\mathrm{exp}}$ are obtained from power-law
fits $\langle x^2\rangle \propto t^{\alpha}$ to the simulated and experimental MSD curves respectively. 
\begin{table}[htbp]
\centering 
\caption{Spatiotemporal scale matching results for all five turtles.
$T_{x,1/2}^{\mathrm{fit}}$: half-period from the \texttt{lmfit} procedure with a sum of exponential and cosine as a fit function. 
$T_{x,1/2}^{\mathrm{raw}}$: half-period from raw VACF minima.
$L_x^{\mathrm{fit}}$ and $L_x^{\mathrm{raw}}$: corresponding characteristic length scales
$\sqrt{\langle x^2(T_{x,1/2})\rangle}$ (km).
$\alpha_{\mathrm{exp}}$, $\alpha_{\mathrm{sim}}$: MSD power-law exponents.}
\label{tab:scale_matching}  

\begin{tabular}{lcccccc}       
\toprule 
\textbf{Turtle} &
  $T_{x,1/2}^{\mathrm{fit}}$ (h) &
  $L_x^{\mathrm{fit}}$ (km) &
  $T_{x,1/2}^{\mathrm{raw}}$ (h) &
  $L_x^{\mathrm{raw}}$ (km) &
  $\alpha_{\mathrm{exp}}$ &
  $\alpha_{\mathrm{sim}}$ \\
\midrule
Mokamba  & 164.8 & 83.9  & 203.6 & 100.2 & 1.68 & 1.57 \\
Catchupa & 161.0 & 66.1  & 163.9 &  67.1 & 1.64 & 1.68 \\
Papaya  & 313.5 & 143.0 & 155.9 &  78.7 & 1.71 & 1.84 \\
Olympia & 138.2 &  73.4 &  95.4 &  55.7 & 1.48 & 1.60 \\
Goody   & 237.4 &  82.4 & 157.2 &  59.8 & 1.55 & 1.64 \\
\bottomrule  
\end{tabular} 
\end{table}

We see that both methods for extracting the half period deliver approximately consistent results for all turtles except Papaya. The spatial length scale identified by this approach matches roughly in order of magnitude to the $x$ extension of the foraging region covered by the turtles. As this is a very handwaving, rough approach one might not expect a more precise matching of scales here. In any case, this argument confirms again the big size of the loops generated intrinsically by the turtles themselves, which appear to be so large that they essentially optimise to cover the chosen foraging region along $x$.

The large discrepancy between $T_{x,1/2}^{\mathrm{fit}}$ and $T_{x,1/2}^{\mathrm{raw}}$ for
Papaya (313.5\,h vs.\ 155.9\,h) reflects the relatively noisy empirical VACF of that turtle.
The \texttt{lmfit} procedure locks onto a longer dominant oscillation period visible in the smoothed fit, 
while the raw-minima detection responds to the first pronounced dip in the unsmoothed signal. 
In the case of Goody, the experimental MSD does not reach a plateau; instead, it appears to exhibit a secondary superdiffusive regime. This corresponds to the extension of the trajectory in the x direction well beyond that predicted by the scale-matching analysis (\textbf{$L_x^{\mathrm{fit}}$ (km)}). Furthermore, Goody's trajectory exhibits relatively larger loops than those of the other turtles, which may have contributed to the observed secondary superdiffusive regime.

\end{document}